\documentclass[a4paper,11pt,fleqn]{article}
\usepackage[utf8]{inputenc}
\usepackage{amsthm}
\usepackage[fleqn]{amsmath}
\usepackage{graphicx}
\usepackage{amssymb}
\usepackage{url}
\usepackage{hyperref}

{\catcode `\@=11 \global\let\AddToReset=\@addtoreset}
\AddToReset{equation}{section}

\author{Jacek Jezierski\thanks{E--mail: \texttt{Jacek.Jezierski@fuw.edu.pl}}\; and Szymon Migacz\thanks{E--mail: \texttt{Szymon.Migacz@gmail.com}} \\
	Department of Mathematical Methods in
	Physics, \\ University of Warsaw,
	ul. Pasteura 5, 02-093 Warszawa, Poland}

\title{Charges of the gravitational field and (3+1) decomposition of CYK tensors part 2}
\newtheorem{theorem}{Theorem}
\newtheorem{lemat}{Lemma}

\begin{document}
\maketitle
%Alternative proof for the fully charged mono-dipole spin-2 field solution,
%asymptotics in asymptotically flat spacetime and asymptotically de Sitter,
%properties of quasi-local charges on the Cauchy surface.
%asymptotyka dla czasoprzestrzeni asymptotycznie płaskich i asymptotycznie de Sitter,
%własności asymptotyczne kwazilokalnych ładunków na pow. danych początkowych.
\begin{abstract}
	
 The work describes the method of construction of charges (conserved quantities) for the gravity field in the (3 + 1) decomposition. The presented construction uses tensors of the electrical and magnetic parts of the Weyl tensor and conformal Killing vectors. In the case of conformally flat spatial hypersurfaces, we get twenty local charges, which  can be expressed in terms of the initial data (three-dimensional metric and extrinsic curvature tensor).
The work shows the relationships between charges which are obtained by this method and the usual ADM approach. In traditional ADM approach Killing vectors are used to construct corresponding charges e.g. time translation corresponds to ADM mass,
spatial translations give linear momentum and rotations correspond to angular momentum.
Gravito-electric and gravito-magnetic charges need conformal Killing vectors e.g. mass corresponds to dilation, linear momentum is related to rotation and angular momentum needs conformal acceleration.
The analyzed example of the Schwarzschild--de Sitter spacetime suggests that in some cases the mass calculated by the presented method has better properties than the traditional ADM mass.
Next we discuss asymptotic charges which are no longer rigidly conserved but rather approach finite value at spatial infinity.
\end{abstract}

%\tableofcontents
%{\small\tableofcontents}
%\listoffigures
%\listoftables
%Szymon, spróbuj przetłumaczyć swoją pracę na angielski odpowiednio ją skracając.
%W szczególności postaraj się odwołać do poprzedniej pracy \cite{CGQ32jjsm} tam gdzie to jest potrzebne.

\section*{Notation and conventions}
%\addcontentsline{toc}{chapter}{Notacja oraz przyjęte konwencje}

\begin{itemize}
	\item $g_{\mu \nu}$ spacetime metric with signature ($-,+,+,+$),
	\item $\gamma_{ i j }$ Riemannian metric induced on three-dimensional
spatial surfaces,
	\item $\eta_{\mu \nu}$ metric of Minkowski spacetime in
		$\mathbb{R}^4$,
	\item $\eta _{AB}$ metric on a two-dimensional sphere, $\eta _{AB}=\textrm{diag}
		\left( r^2, r^2\sin^2\theta \right)$.
\end{itemize}

Indices for which the summation convention applies:
\begin{itemize}
	\item indices written in Greek letters \{$\mu, \nu, \alpha, \beta, \dots$\}
run a set of coordinates of space-time,
	\item indices written in Latin letters \{$a, b, c, d , \dots$\}
		run a set of coordinates on spatial hypersurfaces,
	\item indices written in Latin capital letters \{$A, B, C, \dots$\}
		run a set of angular coordinates on a two-dimensional sphere in
standard parametrization \\ $(\theta,\phi) \in (0,\pi)\times(0,2\pi)$.
\end{itemize}
Indices for which the summation convention does not apply:
\begin{itemize}
	\item \{$x,y,z$\} used for writing in Cartesian coordinates,
	\item \{$r , \theta, \phi$\} used for writing in spherical coordinates.
\end{itemize}
$\nabla$ and $D$ mean four-dimensional and three-dimensional covariant derivatives respectively.
We will also use shortened, symbolic notation in which semicolon ($;$) means a covariant derivative for space-time,
vertical line ($|$) is a covariant derivative on three-dimensional hypersurfaces, and a~double vertical line ($||$) means a covariant derivative on a two-dimensional sphere.

\noindent Symmetrization and antisymmetrization of indices $\alpha,\beta$ we write
respectively as $(\alpha \beta)$ and $[\alpha \beta]$, we assume that both of these
operations contain a numerical factor depending on the number of indices that
include, in particular:
\[
A _{(ij)} = \frac{1}{2}A _{ij} + \frac{1}{2}A _{ji}\,,
\]
\[
A _{[ij]} = \frac{1}{2}A _{ij} - \frac{1}{2}A _{ji}\,,
\]
\[
A _{ij} = A _{(ij)}+A _{[ij]}\,.
\]
For the antisymmetric tensor, we accept the convention:
\[
\sqrt{|g|} \varepsilon^{1 2 \dots n} = 1\,.
\]

%\noindent Skrócony zapis funkcji jako $f(\{a,b\})=\{c,d\}$ należy rozumieć w następujący sposób:
%
%\[
%f(x) = \begin{cases} c & \quad \textrm{dla}\quad x=a \\
%							d & \quad \textrm{dla}\quad x=b \end{cases}
%\]
\noindent In the case of calculation related to the (3 + 1) decomposition, we will write three over tensors specified on the spatial hypersurface $\Sigma$ and four
for objects in four-dimensional spacetime. To simplify the notation,
we omit the index in situations, where it is clearly derived from the context. \\
\noindent To simplify the writing, we omit the integration variables
$\textrm{d}
\theta \textrm{d}\phi$ in the majority of integrals on two-dimensional spheres.\\
\noindent The geometric layout of units was adopted throughout the work $c=G=1$.

\section*{Index of notation}
%\addcontentsline{toc}{chapter}{Indeks oznaczeń}
$g _{\mu\nu}$ -- metric tensor, \\
$\eta_{\mu\nu}$ -- metrics of Minkowski spacetime,\\
$R _{\alpha \beta \mu \nu}$ -- Riemann tensor,\\
$R _{\alpha \beta}$ -- Ricci tensor, \\
$R$ -- scalar of curvature, \\
$W _{\alpha \beta \mu \nu}$ -- Weyl tensor,\\
$E _{\mu \nu}$ -- electric part of Weyl tensor, \\
$B _{\mu \nu}$ -- magnetic part of Weyl tensor, \\
$\Lambda$ -- cosmological constant, \\
$T _{\mu \nu}$ -- energy-momentum tensor, \\
$S(r)$ -- two-dimensional sphere with radius $r$,\\
$\textrm{d}\Omega^2 = r^2\textrm{d}\theta^2 + r^2\sin^2\theta \textrm{d}\phi^2$.\\

The conformal Killing vectors (CKV): \\
$\mathcal{T}_{\mathbf k}$ -- translation generators, \\
$\mathcal{R}_{\mathbf k}$ -- rotation generators,\\
$\mathcal{K}_{\mathbf k}$ -- generators of proper conformal transformations, \\
$\mathcal{S}$ -- scaling generator.\\

(3 + 1) Decomposition: \\
$\Sigma$ --  spatial hypersurface,\\
$N$ -- lapse function, \\
$N^k$ --  shift vector, \\
$\gamma _{ij}$ -- Riemann metric on a hypersurface $\Sigma$,\\
$K _{ij}$ -- tensor of the extrinsic curvature, \\
$K$ -- trace of the tensor of the extrinsic curvature $K=K _{ij}g ^{ij}$,\\
$P _{ij}$ -- canonical  ADM momentum $P _{ij} = K _{ij} - \gamma _{ij}K$,\\
$n^k$ -- normalized normal vector, \\
$	(A \wedge B)_a :=
	\varepsilon _{a}{} ^{bc}A _{b}{}^dB _{dc}$.\\

%%%%%%%%%%%%%%%%%%%%%%%%%%%%%%%%%%%%%%%%%%%%%%%%%%%%%%%%%%%%%%%%%%%%%%%%%%%%%%%%
%                                 WPROWADZENiE                                 %
%%%%%%%%%%%%%%%%%%%%%%%%%%%%%%%%%%%%%%%%%%%%%%%%%%%%%%%%%%%%%%%%%%%%%%%%%%%%%%%%

%%%%%%%%%%%%%%%%%%%%%%%%%%%%%%%%%%%%%%%%%%%%%%%%%%%%%%%%%%%%%%%%%%%%%%%%%%%%%%%%
%                              WEYL introduction                               %
%%%%%%%%%%%%%%%%%%%%%%%%%%%%%%%%%%%%%%%%%%%%%%%%%%%%%%%%%%%%%%%%%%%%%%%%%%%%%%%%

%\section*{Wprowadzenie}
%\addcontentsline{toc}{chapter}{Wprowadzenie}
\section{Introduction}

The aim of this work is to provide definitions and discuss basic properties of
gravitational charges, or quantities specified on the spatial hypersurface
 $\Sigma$ immersed in four-dimensional spacetime.
The structure of quasi-local charges presented here can be used for
any initial data $(\gamma _{ij}, K _{ij})$, for which
three-dimensional metric $\gamma _{ij}$ is conformally flat %(or asymptotically flat)
or for the three-metric which approaches conformally flat metric at infinity.

\noindent The idea of defining charges as a contraction of the electrical and magnetic parts of
the Weyl tensor with conformal Killing vectors results from the structure
of conformal Yano--Killing tensors (CYK tensors) and spin-2 field in linearized gravity.
 In \cite{SM} it has been shown that
the contractions of the spin-2 field with CYK tensors give  closed two-forms (satisfy the Gaussian law), %represented by a closed two-form,
 and therefore we can consider them as charges.  The four-dimensional Minkowski spacetime has
twenty linearly independent CYK tensors, therefore we get twenty independent
charges. Unfortunately, this method of constructing charges can not be used for
 Schwarzschild spacetime, where the equation
defining CYK tensors has only two independent solutions.\\
In \cite{SM} has been proved the
lemma which says that each of the twenty basic CYK tensors in Minkowski space
 can be decomposed into a sum of expressions containing a function depending on time $t$ and
 the product of the field $\partial_t$ with the conformal Killing vector (plus the dual field
  to such a product). The lemma suggests that in the case of spacetime with not
enough CYK tensors, we can try to define charges with help of
conformal Killing vectors, which is the main goal of this work.

%\vskip 3mm
\noindent Section \ref{chapterLinearized} contains a short overview of
linearized gravity as a spin-2 field theory. We present formulae
defining charges in the linearized case and we calculate their values for
``a fully charged solution'' derived from \cite{1}.
%\vskip 3mm

\noindent In section \ref{ladunkiKwazilokalne} we define  twenty
charges for the gravitational field, using
electrical and magnetic parts of spin-2 field, and conformal Killing vectors. We show how
the initial data $(\gamma_{ij}, K_{ij})$ on the hypersurface $\Sigma$ enables one to recover
the electrical and magnetic parts of the Weyl tensor for vacuum spacetime (with a cosmological constant).
We also formulate conditions guaranteeing independence of charges from the choice of
integration surface, or analog of Gaussian electromagnetic law for gravity.
The relationship between ``gravito-electromagnetic'' charges with
linear and angular momentum is presented.
%\vskip 3mm

\noindent Section \ref{chapterSchwarzschild} is devoted to
Schwarzschild--de Sitter spacetime.
We analyze two families of foliations with spatial hypersurfaces.
The first one consists of the
 usual constant Schwarzschildean time slices, characterized by zero extrinsic curvature
 and conformally flat internal geometry. The second one is a set of
surfaces with non-zero extrinsic curvature, equipped with a flat
Euclidean metric. In both cases, we compare the ``gravito-electromagnetic'' mass
with the ADM mass.
%\vskip 3mm

\noindent In section \ref{chapterAsymptotyczne} we prove the claim
allowing to define asymptotic charges for a particular class of
 asymptotically flat spacetimes.

%\vskip 3mm

%\noindent The rest of this section contains definitions and basic properties of
%objects used at work.
%\vskip 3mm

\subsection{Electrical and magnetic parts of Weyl tensor}
\label{introWeyl}
Four-dimensional Weyl tensor has ten independent components.
It is expressed by  Riemann tensor and Ricci tensor by the following formula:
\begin{equation}
\label{defweyl}
	W _{\alpha \beta \mu \nu} = R _{\alpha \beta \mu \nu}-(g _{\alpha [ \mu}R
	_{\nu]\beta}- g _{\beta[\mu}R _{\nu]\alpha}) + \frac{1}{3}R g _{\alpha [\mu}g
	_{\nu]\beta}
\end{equation}
and has the following properties:
\begin{itemize}
	\item symmetry of exchange of index pairs: $W _{\alpha\beta\mu\nu}=W
		_{\mu\nu\alpha\beta}$,
	\item antisymmetry in the first pair of indices: $W _{\alpha\beta\mu\nu}=-W
		_{\beta\alpha\mu\nu}$,
	\item antisymmetry in the second pair of indices: $W _{\alpha\beta\mu\nu}= -W
		_{\alpha\beta\nu\mu}$,
	\item tracelessness: $W ^{\alpha}{}_{\beta \alpha \nu}=0$,
	\item vanishing antisymmetrization of the last three indices: $W _{\alpha
		[\beta\mu\nu]}=0$.
\end{itemize}
Symmetries of  Weyl tensor enable us to define the following dual tensors:
\begin{equation}
	^*W _{\alpha\beta\mu\nu}:=\frac{1}{2}\varepsilon_{\alpha\beta}{}^{\lambda \delta}
	W _{\lambda \delta \mu \nu}\,, \quad\quad\quad
	W^*{} _{\alpha\beta\mu\nu}:=\frac{1}{2} W _{\alpha\beta\lambda\delta}\varepsilon
	^{\lambda\delta}{}_{\mu\nu}\,.
\end{equation}
Consider a spatial hypersurface $\Sigma$ immersed in a four-dimensional
pseudoriemannian manifold $(\mathcal{M},g)$, let us denote by $n$
unit normal vector to $\Sigma$. Using $(3+1)$ decomposition the Weyl tensor can be split into gravito-electric and gravito-magnetic parts:
\begin{equation}
	E _{\alpha\beta}:= W _{\alpha\mu\nu\beta}n^\mu n^\nu\,,
\end{equation}
\begin{equation}
	B _{\alpha\beta}:= W^*{} _{\alpha\mu\nu\beta}n^\mu n^\nu\,.
\end{equation}
Tensors $E$ and $B$  are symmetric, traceless and spatial (it follows
directly from the  property of Weyl tensor) and both have five independent
components.

%%%%%%%%%%%%%%%%%%%%%%%%%%%%%%%%%%%%%%%%%%%%%%%%%%%%%%%%%%%%%%%%%%%%%%%%%%%%%%%%
%                                Pola Killinga                                 %
%%%%%%%%%%%%%%%%%%%%%%%%%%%%%%%%%%%%%%%%%%%%%%%%%%%%%%%%%%%%%%%%%%%%%%%%%%%%%%%%

\subsection{Killing fields and their conformal generalization}
\label{introCKV}
The Killing field is a vector field defined on the Riemannian manifold
(or pseudoriemannian) which preserves the metric. We can treat it as
infinitesimal isometry generator (flow generated by Killing field
 is an isometry of the manifold). By definition, a vector field $X$ is a Killing field (KV) if it satisfies the equation:
\begin{equation}
	\mathcal{L}_X g = 0 \,,
\end{equation}
where $g$ is a metric tensor, and $\mathcal{L}$ is the Lie derivative.\\
If we restrict ourselves to the Levi-Civita connection, the above equation becomes equivalent
to:
\begin{equation}
	\label{killinga}
	\nabla_{(i}X _{ j)}=0\,.
%	X _{(i | j)}=0\,.
\end{equation}
The flat three-dimensional Euclidean space has six linearly independent
Killing fields: three fields corresponding to translation generators $\mathcal{T}_{\mathbf k}$ and three
corresponding to rotation generators $\mathcal{R}_{\mathbf k}$. In appendix (\ref{KVflatspace})
a complete solution to the equation  (\ref{killinga}) for
$(\mathbb{R}^3,\eta)$ is presented.

\noindent Generally, Killing fields are not preserved by conformal transformations. If we assume that the vector field $X$ is a Killing vector on a manifold
$(\mathcal{M},g)$,
then after conformal transformation into manifold
$(\mathcal{M},\tilde{g})$, where $\tilde{g} _{ij} = e^{2\Omega} g _{ij}$,
 we can receive $\tilde{g} _{k(i} \tilde\nabla_{j)} X^k \neq 0$.

\noindent Let us generalize Killing fields so that the new vectors will be
preserved by conformal transformations, to this end
let us modify the equation (\ref{killinga}) replacing zero on the right hand side by
expression proportional to the metric:
\begin{equation}
	\label{konfkillinga}
	\nabla_j X _{i } + \nabla_i X _{j} = \lambda g _{i j} \,,
\end{equation}
where $\lambda$ is arbitrary function which is related to the conformal factor. \\
Vector field $X$ which satisfies the equation (\ref{konfkillinga}) is called
conformal Killing field (CKV). We see that every Killing field is
simultaneously a conformal Killing field (for $\lambda=0$). In general case
function $\lambda$ depends on the field $X$, and let $n$ be a dimension
of our manifold. Trace of the equation (\ref{konfkillinga}) yields:
\begin{equation}
	\label{ckillambda}
	2X^i{}_{;i} = n \lambda \quad\Rightarrow\quad \lambda =
	\frac{2}{n}X^i{}_{;i}\,.
\end{equation}
So-defined vector fields remain invariant under the influence of
conformal transformations.
\begin{lemat}
	\label{CKVflatlemma}
If $X^a$ is a conformal Killing field for a metric $g _{ij}$ and
conformal factor $\lambda$, than is also a conformal Killing field for the metric
$\tilde{g} _{ij}= e^{2\Omega} g _{ij}$ and conformal factor $\tilde{\lambda}
= \lambda + 2\Omega _{,a}X^a$.
\end{lemat}
\noindent \textit{Proof:}\\
If $\tilde{g} _{ij} = e^{2\Omega} g _{ij}$ than the following
transformation law for Christoffel symbols is true \cite{wald}:
\begin{equation}
	\tilde{\Gamma} ^{k}{}_{ij} = \Gamma ^{k}{}_{ij}  + \delta
	^{k}{}_{i}\partial_j \Omega + \delta ^{k}{}_{j} \partial_i \Omega - g
	_{ij}\nabla^k \Omega\,,
\end{equation}
where $\tilde{\Gamma} ^{k}{}_{ij}$ is the Christoffel symbol for the metric
$\tilde{g}$, and $\Gamma ^{k}{}_{ij}$ for the metric $g$.\\
Covariant derivative transforms under conformal rescaling in the following way:
\begin{equation}
\label{ckvconformal}
	\begin{aligned}
	\tilde{g}_{kj}\tilde\nabla _{i}X^k =& \tilde{g} _{kj} (X ^{k}{}_{,i}+\tilde\Gamma
	^{k}{}_{mi} X^m) = \tilde{g}_{kj} \left[X ^{k}{}_{,i}+X^m(\Gamma
	^{k}{}_{mi}+\delta ^{k}_{m}\partial_i \Omega + \delta ^{k}_{i} \partial_m
	\Omega - g _{mi}\nabla^k\Omega)\right]=\\
	=&\tilde{g} _{kj} \left[ X ^{k}{}_{,i} +X^m \Gamma ^{k}{}_{mi} \right]
	+ e^{2\Omega} X _{j}\partial_i \Omega + \tilde{g} _{ij}X^m\partial_m \Omega -
	e^{2\Omega} X _{i}\partial_j\Omega = \\
	=&\tilde{g} _{kj} \nabla_i X^k + \tilde{g} _{ij}X^m\partial_m \Omega
	+2e^{2\Omega}X _{[i}\partial _{j]}\Omega\,.
	\end{aligned}
\end{equation}
Assuming that $X^k$ is a conformal Killing vector for $g$:
\begin{equation}
	g_{kj}\nabla_i X^k + g_{ki}\nabla_j X^k = \lambda g_{ij} \quad \Rightarrow \quad
	\tilde{g}_{kj} \nabla_i X^k + \tilde{g} _{ki}\nabla_j X^k = \lambda \tilde{g}_{ij}\,.
\end{equation}
 Symmetrizing (\ref{ckvconformal}), we obtain the equation for conformal Killing vectors:
\begin{equation}
	2\tilde{g}_{k(j}\tilde\nabla_{i)} X^k = \tilde{g}_{ij}(\lambda + 2 X^m\nabla_m
	\Omega)\,.
\end{equation}
 \qed \\
%\vskip 3mm
\noindent In a flat (according to the thesis of lemma, also in a conformally flat) three-dimensional space, we have ten linearly independent conformal Killing fields.
 Six
of them are previously introduced fields of translation generators $\mathcal{T}_{\mathbf k}$ and rotation generators
$\mathcal{R}_{\mathbf k}$, the other four fields correspond to the scaling $\mathcal{S}$ and
three ``appropriate'' conformal transformations $\mathcal{K}_{\mathbf k}$. In the Cartesian coordinate system, expressions for the fields take the following form:
\begin{equation}
	\label{CKVT}
	\mathcal{T}_{\mathbf k}=\frac{\partial}{\partial x^k}\,,
\end{equation}
\begin{equation}
	\label{CKVR}
	\mathcal{R}_{\mathbf k}=\varepsilon_k{}^{ij}x_i\frac{\partial}{\partial x^j}\,,
\end{equation}
\begin{equation}
	\label{CKVK}
	\mathcal{K}_{\mathbf k}=x_k\mathcal{S}-\frac{1}{2}r^2\frac{\partial}{\partial x^k}\,,
\end{equation}
\begin{equation}
	\label{CKVS}
	\mathcal{S}=x^k \frac{\partial}{\partial x^k}\,,
\end{equation}
where $r^2=x^2+y^2+z^2$. %\\
The full calculations leading to the above formulae are included in the appendix
\ref{CKVflatspace}. \\
Note that the above definitions in a natural way distinguish one point --- the center of the coordinate system. \\
It means that the above conformal Killing fields behave with respect to active translations.
Consider two Cartesian coordinate systems $x$, $y$ shifted by constant
vector $a$ so that the relationship $y^k = x^k + a^k$ is fulfilled. Let us denote by
$\mathcal{A}(x)$ a vector field $\mathcal{A}$ related to the coordinate system
$x$. Under the influence of translation by the vector $a^k$, Killing fields are transformed in the following way:
\begin{equation}
	\label{transformationfirst}
\mathcal{T}_k(x)=\mathcal{T}_k(y)\,,
\end{equation}
\begin{equation}
\mathcal{S}(y) = \mathcal{S}(x)+\mathcal{T}_ka^k\,,
\end{equation}
\begin{equation}
\mathcal{R}_k(y)=\mathcal{R}_k(x)+\varepsilon_k{}^{ij}a_i\mathcal{T}_j(x)\,,
\end{equation}
\begin{equation}
	\label{transformationlast}
\mathcal{K}_k(y)=\mathcal{K}_k(x)+a_k\mathcal{S}(x)+a^l\mathcal{R}_{kl}(x)+\mathcal{T}_l(x)(a_ka^l-\frac{1}{2}a^ia_i\delta^l_k)\,,
\end{equation}
where:
\[
\mathcal{R}_{kl}:=x_k\partial_l-x_l\partial_k\,.
\]

\subsection{(3 + 1) Decomposition of Einstein equations}
Consider a four-dimensional space-time $(\mathcal{M},g)$ which is a solution of Einstein equations (without cosmological constant):
\begin{equation}
	R _{\mu\nu}-\frac{1}{2}R g _{\mu \nu}=8\pi T _{\mu\nu}\,.
\end{equation}
We assume that spacetime $(\mathcal{M}, g)$ is globally
hyperbolic, which means that there is a spatial hypersurface $\Sigma$,
whose intersection with any timelike or null curve is exactly one
point. By $\Sigma_t$ we denote foliation of $\mathcal{M}$
by spatial surfaces. Einstein equations in the (3 + 1) decomposition are obtained by projecting onto $\Sigma_t$ and on the direction perpendicular to
$\Sigma_t$. \\
Let us denote by $n$ the unit normal vector to the surface $\Sigma_t$ and by $\gamma$
the projection operator at~$\Sigma_t$. Einstein equations can be written in
an equivalent form which is more convenient for projection:
\begin{equation}
	\label{equivEinstein}
	R _{\mu \nu} = 8\pi \left( T _{\mu \nu} -\frac{1}{2}T g _{\mu\nu} \right)\,,
\end{equation}
where $T$ is the trace of the energy-momentum tensor, $T:=g ^{\mu\nu}T _{\mu\nu}$.\\
The vector $n$ is timelike, so it can be treated as
four-speed of an observer.
The world lines of such observer are perpendicular to
hypersurface $\Sigma_t$, this means that the surfaces $\Sigma_t$ locally represent simultaneous events from the point of view of the observer.
Let us define:
\begin{equation}
	{\cal E}:=T _{\mu\nu}n^\nu n^\mu
\end{equation}
and:
\begin{equation}
	p_\alpha := -T _{\mu\nu}n^\mu \gamma ^{\nu}{}_{\alpha}\,.
\end{equation}
${\cal E}$ and $p_\alpha$ have respectively an interpretation of the energy density and the momentum density measured by the observer. We can introduce in the same way:
\begin{equation}
	S _{\alpha\beta}:= T _{\mu\nu}\gamma ^{\mu}{}_{\alpha}\gamma ^{\nu}{}_{\beta}
\end{equation}
and the trace:
\begin{equation}
	S:= \gamma ^{ij}S _{ij}= g ^{\mu\nu} S _{\mu\nu}\,.
\end{equation}
Applying the projection operator $\gamma^{\mu}{_\nu}$ to the Einstein equations (\ref{equivEinstein}) and using Gaussian formula (\ref{gaussRelation}) we obtain: %($m:=Nn$):
\begin{equation}
	\mathcal{L}_n K _{\alpha\beta}=-D_\alpha D_\beta N + N \left[ R
	_{\alpha\beta}+K K _{\alpha\beta}-2K _{\alpha\mu}K ^{\mu}{}_{\beta}+4\pi
	\left( (S-{\cal E})\gamma _{\alpha\beta}-2S _{\alpha\beta} \right) \right]\,.
\end{equation}
where $\mathcal{L}_n$ is the Lie derivative with respect to the timelike unit normal $n$.
Note that in the above equation, each of the tensors is tangent to $\Sigma_t$,
we can limit ourselves to spatial indices:
\begin{equation}
	\mathcal{L}_n K _{ij}=-D_i D_j N + N \left[ R
	_{ij}+K K _{ij}-2K _{ik}K ^{k}{}_{j}+4\pi
	\left( (S-{\cal E})\gamma _{ij}-2S _{ij} \right) \right]\,.
\end{equation}
Projection of the Einstein equations (\ref{equivEinstein}) on the direction perpendicular to $\Sigma_t$ gives\footnote{Let us recall that $n$ is normalized such that $g _{\mu\nu}n^\mu n^\nu=-1$.}:
% which amounts to applying the Einstein equations for a pair of vectors $(\mathbf{n},
%\mathbf{n})$.
%As before, we assume the normalization of $n$, i.e.
%$g _{\mu\nu}n^\mu n^\nu=-1$
\begin{equation}
	R _{\mu \nu}n^\mu n^\nu + \frac{1}{2}R = 8\pi T _{\mu\nu}n^\mu n^\nu\,.
\end{equation}
Using the contracted Gaussian formula (\ref{contractedGauss}) and the definition of ${\cal E}:= T _{\mu\nu}n^\mu n^\nu$ yields to the the scalar constraint:
\begin{equation}
	R + K^2 - K _{ij}K ^{ij}=16\pi {\cal E}	\,.
\end{equation}
Projecting Einstein equations (\ref{equivEinstein}) once in $\Sigma_t$ and the second time in the normal direction $n$  leads to the equation:
\begin{equation}
	\overset{4}{R} _{\mu\nu}n^\mu \gamma
	^{\nu}{}_{\alpha}-\frac{1}{2}\overset{4}{R} g _{\mu\nu}n^\mu \gamma
	^{\nu}{}_{\alpha} = 8\pi T _{\mu\nu}n^\mu \gamma ^{\nu}{}_{\alpha}\,.
\end{equation}
Noting that $g _{\mu\nu}n^\mu \gamma ^{\nu}{}_{\alpha}=0$, with the help of contracted Codazzi formula (\ref{contractedcodazzi}) we obtain:
\begin{equation}
	D_j K ^{j}{}_{i} - D _{i}K = 8\pi p_i\,.
\end{equation}
The above equation is called the vector constraint.

%%%%%%%%%%%%%%%%%%%%%%%%%%%%%%%%%%%%%%%%%%%%%%%%%%%%%%%%%%%%%%%%%%%%%%%%%%%%%%%%
%                           ZLINEARYZOWANA GRAWITACJA                           %
%%%%%%%%%%%%%%%%%%%%%%%%%%%%%%%%%%%%%%%%%%%%%%%%%%%%%%%%%%%%%%%%%%%%%%%%%%%%%%%%

\section{Linearized gravity}
\label{chapterLinearized}

%%%%%%%%%%%%%%%%%%%%%%%%%%%%%%%%%%%%%%%%%%%%%%%%%%%%%%%%%%%%%%%%%%%%%%%%%%%%%%%%
%                       Elektr. i magn. pola o spinie 2                        %
%%%%%%%%%%%%%%%%%%%%%%%%%%%%%%%%%%%%%%%%%%%%%%%%%%%%%%%%%%%%%%%%%%%%%%%%%%%%%%%%

\subsection{The electrical and magnetic part of the spin-2 field}
Consider Minkowski space (in the Cartesian coordinate system) equipped with an additional structure --- the field $W _{\alpha\beta\mu\nu}$ which we call the spin-2 field. The field $W$ can be interpreted as a Weyl tensor for linearized gravity, which is defined by the following identities:
\begin{equation}
	W_{\alpha\beta\mu\nu}=W _{\mu \nu \alpha\beta}=W _{[\alpha\beta][\mu\nu]}\,,
\end{equation}
\begin{equation}
	W _{\alpha [\beta\mu\nu]}=0, \quad \quad \eta ^{\alpha\mu}W
	_{\alpha\beta\mu\nu}=0\,.
\end{equation}
We introduce dual fields due to the first and second pair of indices:
\begin{equation}
	^*W _{\alpha\beta\mu\nu}:=\frac{1}{2} \varepsilon _{\alpha\beta\lambda\delta}W
	^{\lambda\delta}{}_{\mu\nu}\,,
\end{equation}
\begin{equation}
	W{}^* {}_{\alpha\beta\mu\nu}:=\frac{1}{2} W _{\alpha\beta}{}^{\lambda\delta}
	\varepsilon _{\lambda\delta \mu \nu}\,,
\end{equation}
which have the following properties:
\begin{equation}
	^*W^* {} _{\alpha\beta\mu\nu} = \frac{1}{4} \varepsilon
	_{\alpha\beta\lambda\delta}W ^{\lambda\delta\rho\sigma} \varepsilon
	_{\rho\sigma\mu\nu}, \quad ^*W = W^*, \quad ^*(^*W) = ^*W^* = -W\,.
\end{equation}
Bianchi identities act as field equations:
\begin{equation}
	\label{fieldequation}
	\nabla _{[\lambda}W _{\mu\nu]\alpha\beta}=0\,,
\end{equation}
which can be formulated equivalently as:
\begin{equation}\label{Bianchiidentities}
	\nabla ^\mu	W _{\mu\nu\alpha\beta}=0 \quad\textrm{or}\quad \nabla _{[\lambda}{}^*W
	_{\mu\nu]\alpha\beta}=0 \quad\textrm{or}\quad \nabla^\mu {}^*W
	_{\mu\nu\alpha\beta}=0\,.
\end{equation}
Weyl tensor can be splitted into the electrical part $E$ and the magnetic part $B$:
\begin{equation}
	\label{linearE}
E_{kl}:=W_{k00l}\,,
\end{equation}
\begin{equation}
	\label{linearB}
B_{kl}:=\frac{1}{2} W_{0kij}\varepsilon^{ij}{}_l\,.
\end{equation}
Electrical (magnetic) part of Weyl tensor we shortly call gravito-electric (gravito-magnetic) tensor.
The electrical and magnetic parts fulfill the following dynamic equations resulting from field equations (\ref{fieldequation}):
\begin{equation}
	\label{dynE}
	\dot{E}_{ij}=\varepsilon_j{}^{lm}\partial_l B_{im}\,,
\end{equation}
\begin{equation}
	\label{dynH}
	\dot{B}_{ij}=-\varepsilon_j{}^{lm}\partial_l E_{im}
\end{equation}
and algebraic constraint equations:
\begin{equation}\label{constrE}
	E^i{}_i=0\,,
\end{equation}
\begin{equation}\label{constrB}
	B^i{}_i=0\,.
\end{equation}
There are also differential constraints:
\begin{eqnarray}
    { E}^k{_{l|k}}  &=&
    %{ E}^k{_{l|k}} - K^{ij}{ W}^0{_{ijl}}
   \nabla_k { W}^{k00}{_l} = \nabla_\mu { W}^{\mu 00}{_l} = 0 \, , \label{divDs2}\\
  %{\cal W}^{0k}{_{ij|k}} + K_{kj}{\cal D}^k{_i} - K_{ki}{\cal D}^k{_j} &=& \nabla_k {\cal W}^{0k}{_{ij}} %= \nabla_\mu {\cal W}^{0\mu}{_{ij}} = 0 \, , \label{divBs2}\\
   { B}^{kl}{_{|k}}  &=&  \frac12\varepsilon^{ijl} \nabla_\mu { W}^{0\mu}{_{ij}} = 0
    \, , \label{divBs3} %\\ %\nonumber
%  0=N\nabla_{\mu} {\cal W}^{0k\mu}{_l} &=& \partial_0 (N{\cal W}^{0k0}{_l}) + D^j( N{\cal W}^{0k}{_{jl}})
%  -D_i (N^i N {\cal W}^{0k0}{_l}) \\ \nonumber
%  & & + 2N^2K_{jl} {\cal W}^{0k0j}+ K^{ij}{\tilde g}^{km} {\cal W}_{mjil}
%   - N^2K^k{_j} {\cal W}^{0j0}{_l} \\ & &
%   + D^j N {\tilde g}^{km} {\cal W}_{jm}{^0}{_l}  +
%  ND_j N^k {\cal W}^{0j0}{_l} -N D_l N^j {\cal W}^{0k0}{_j}
%   \\
%  0=N\nabla_{\mu} {\cal W}^{0k\mu}{_l} &=& \nonumber
%  \partial_0 ({\cal D}^{k}{_l}) + D^j( N{\cal B}^{km}\varepsilon{_{jlm}})
%  -D_i (N^i {\cal D}^{k}{_l}) \\ \nonumber
%  & & + 2NK_{jl} {\cal D}^{kj} +NK^{ij}\varepsilon^k{_{jm}}\varepsilon_{lin} {\cal D}^{mn}
%   - NK^k{_j} {\cal D}^{j}{_l} \\ & &
%   + D^j N \varepsilon^k{_{ij}} {\cal B}^i{_l}  +
%  D_j N^k {\cal D}^{j}{_l} - D_l N^j {\cal D}^{k}{_j} \, , \label{rotHs2}
% \\
%  0=N\nabla_{\mu} {^*}\!{\cal W}^{0k\mu}{_l} &=& \nonumber
%  \partial_0 ({\cal B}^{k}{_l}) - D^j( N{\cal D}^{km}\varepsilon{_{jlm}})
%  -D_i (N^i {\cal B}^{k}{_l}) \\ \nonumber
%  & & + 2NK_{jl} {\cal B}^{kj} +NK^{ij}\varepsilon^k{_{jm}}\varepsilon_{lin} {\cal B}^{mn}
%   - NK^k{_j} {\cal B}^{j}{_l} \\ & &
%   - D^j N \varepsilon^k{_{ij}} {\cal D}^i{_l}  +
%  D_j N^k {\cal B}^{j}{_l} - D_l N^j {\cal B}^{k}{_j}
%   \, . \label{rotEs2}
\end{eqnarray}
which can be deduced from (\ref{Bianchiidentities}).
Correctness of the formulae (\ref{dynE}) and (\ref{dynH}) can be easily verified. Using the equations of $E_{kl}$ and $B_{kl}$ for the equation (\ref{dynE}) we obtain a part of Bianchi identities (\ref{Bianchiidentities}):
\begin{equation}
	\begin{aligned}
		\dot{E}_{ij}-\varepsilon_{j}{}^{lm}\partial_l B_{im}&=\partial_0
		 W_{i00j}-\varepsilon_{j}{}^{lm}\partial_l(\frac{1}{2}W_{0iab}\varepsilon^{ab}{}_m)&=& \\
		&=\partial_0 W_{i0j}{}^0-\frac{1}{2}\partial_l
		W_{0iab}(\delta^a_j\delta^{lb}-\delta_j^b\delta^{la})&=&\\
		&=\partial_0 W_{i0j}{}^0-\frac{1}{2}\partial_l W_{0ij}{}^l+\frac{1}{2}\partial_l
		W_{0i}{}^l{}_j &=&\\
		&=\partial_\mu W_{i0j}{}^\mu &=&\quad 0\,.
\end{aligned}
\end{equation}
Analogically for the equation (\ref{dynH}):
\begin{equation}
	\begin{aligned}
		\dot{B}_{ij}+\varepsilon_j{}^{lm}\partial_l
		 E_{im}&=\partial_0(\frac{1}{2}W_{0iab}\varepsilon^{ab}{}_j)-\varepsilon_j{}^{lm}\partial_l
		W_{0i0m}&=&\\
		&=\varepsilon_j{}^{lm}(\frac{1}{2}\partial_0W_{0ilm}-\partial_lW_{0i0m})&=&\\
		&=\varepsilon_j{}^{lm}(\frac{1}{2}\partial_lW_{m00i}+\frac{1}{2}\partial_m
		W_{l00i})&=&\quad 0\,.
	\end{aligned}
\end{equation}
Compatibility of dynamic equations (\ref{dynE}) and (\ref{dynH}) with the constraints (\ref{constrE}) and (\ref{constrB}):
\begin{equation}
\dot{E}_i{}^i=\varepsilon^{ilm}\partial_lB_{im}=0\,,
\end{equation}
\begin{equation}
\dot{B}_i{}^i=-\varepsilon^{ilm}\partial_lE_{im}=0\,.
\end{equation}

\subsection{Definition of charges}
By charge we mean an integral of a contraction of electrical or magnetic part of spin-2 field with a conformal Killing vector:
\begin{equation}
	\label{chargesE}
	Q(E,X):=\int _{S(r)}E _{i}{}^{j} X^i dS_j \,, %=\int_{S(r)}\{E_{ij}, H_{ij}\}X^in^jdS
\end{equation}
\begin{equation}
	\label{chargesB}
	Q(B,X):=\int _{S(r)} B _{i}{}^{j} X^i dS_j \,. %=\int_{S(r)}\{E_{ij}, H_{ij}\}X^in^jdS
\end{equation}
For three-dimensional Euclidean space, the vector space of conformal Killing fields is ten-dimensional. It allows us to define twenty independent charges in that way.
Let us examine the dependence of charges on time. Applying a differentiation over time (marked with a dot) to (\ref{chargesE}) and (\ref{chargesB}) we receive:
\begin{equation}
	\label{chargesEtimederivative}
	\dot{Q}(E,X)=\int _{S(r)}\dot{E} _{i}{}^{j} X^i dS_j\,,
\end{equation}
\begin{equation}
	\label{chargesBtimederivative}
	\dot{Q}(B,X)=\int _{S(r)} \dot{B}_{i}{}^{j} X^i dS_j\,.
\end{equation}
Next, use the dynamic equations (\ref{dynE}), (\ref{dynH}) for the electrical and magnetic part:
\begin{equation}
	\label{dotEX}
	\dot{Q}(E,X)=\int_{S(r)} \varepsilon ^{jlm}\partial_l B _{im}X^idS_j\,,
\end{equation}
\begin{equation}
	\label{dotHX}
	\dot{Q}(B,X)=\int_{S(r)} -\varepsilon ^{jlm}\partial_l E _{im}X^idS_j\,.
\end{equation}
Integrating by parts:
\begin{equation}
	\label{dotEXparts}
	\dot{Q}(E,X)=\int_{S(r)} -\varepsilon ^{jlm} B _{im}\partial_lX^idS_j\,,
\end{equation}
\begin{equation}
	\label{dotHXparts}
	\dot{Q}(B,X)=\int_{S(r)} \varepsilon ^{jlm} E _{im}\partial_lX^idS_j\,.
\end{equation}
The spatial derivatives of the basic conformal Killing vectors gives:
\begin{equation}
	\label{partialT}
	\partial_j\mathcal{T}^i_k=0\,,
\end{equation}
\begin{equation}
	\label{partialS}
	\partial_j\mathcal{S}^i=\delta^i_j\,,
\end{equation}
\begin{equation}
	\label{partialR}
	\partial_j\mathcal{R}^i_k = \partial_j(\varepsilon_k{}^{li}x_l)=
	\varepsilon_{kj}{}^i\,,
\end{equation}
\begin{equation}
	\label{partialK}
	\partial_j\mathcal{K}^i_k=\partial_j(x_kx^i-\frac{1}{2}r^2\delta^i_k)=
	\delta_{jk}x^i+x_k\delta^i_j-x_j\delta^i_k\,.
\end{equation}
Using the results (\ref{partialT})--(\ref{partialK})  we can simplify the
expressions for the time derivatives of charges (\ref{dotEXparts}) and (\ref{dotHXparts}):
\begin{equation}
	\dot{Q}(E, \mathcal{T}_k) = 0 \,,
\end{equation}
\begin{equation}
	\dot{Q}(E, \mathcal{S}) = -\int _{S(r)} \varepsilon ^{jlm}B _{im}\delta
	^{i}{}_{l} dS_j = 0 \,,
\end{equation}
\begin{equation}
	\begin{aligned}
	\dot{Q}(E,\mathcal{R}_k)& = -\int _{S(r)}\varepsilon ^{jlm}B _{im}\varepsilon
		_{kl}{}^i dS_j = -\int _{S(r)}\varepsilon ^{mjl}\varepsilon_l{}^i{}_k B
		_{im}dS_j&=&\\
	& =- \int _{S(r)}(\delta ^{mi}\delta ^{j}{}_{k} - \delta ^{m}{}_{k} \delta
		^{ji}) B _{im}dS_j &=& \\
	&= \int _{S(r)} B ^{j}{}_{k} dS_j =  \int _{S(r)}B_i{}^j
		\mathcal{T}_k^idS_j &=& \; Q(B, \mathcal{T}_k) \,,
	\end{aligned}
\end{equation}
\begin{equation}
	\begin{aligned}
	\dot{Q}(E, \mathcal{K}_k) & = -r\int _{S(r)}\varepsilon ^{jlm}B
		_{im}(\delta _{lk}n^i+n_k\delta ^{i}{}_{l}-n_l \delta ^{i}{}_{k})n_j dS
		&=& \\
	& = -r\int _{S(r)} \varepsilon _{k}{}^{mj} B _{im}n^in_j dS = \int _{S(r)} B
		_{mi}\varepsilon _k{}^{jm} x_j n^i dS &=&\; Q(B,\mathcal{R}_k) \,,
	\end{aligned}
\end{equation}

\begin{equation}
	\dot{Q}(B, \mathcal{T}_k) = 0 \,,
\end{equation}
\begin{equation}
	\dot{Q}(B, \mathcal{S}) = \int _{S(r)} \varepsilon ^{jlm}E _{im}\delta
	^{i}{}_{l} dS_j = 0 \,,
\end{equation}
\begin{equation}
	\begin{aligned}
	\dot{Q}(B,\mathcal{R}_k)& = \int _{S(r)}\varepsilon ^{jlm}E _{im}\varepsilon
		_{kl}{}^i dS_j = \int _{S(r)}\varepsilon ^{mjl}\varepsilon_l{}^i{}_k E
		_{im}dS_j&=&\\
	& = \int _{S(r)}(\delta ^{mi}\delta ^{j}{}_{k} - \delta ^{m}{}_{k} \delta
		^{ji}) E _{im}dS_j &=& \\
	&=- \int _{S(r)} E ^{j}{}_{k} dS_j = - \int _{S(r)}E_i{}^j
		\mathcal{T}_k^idS_j &=&\; -Q(E, \mathcal{T}_k) \,,
	\end{aligned}
\end{equation}
\begin{equation}
	\begin{aligned}
	\dot{Q}(B, \mathcal{K}_k) & = r\int _{S(r)}\varepsilon ^{jlm}E
		_{im}(\delta _{lk}n^i+n_k\delta ^{i}{}_{l}-n_l \delta ^{i}{}_{k})n_j dS
		&=& \\
	& = r\int _{S(r)} \varepsilon _{k}{}^{mj} E _{im}n^in_j dS = -\int _{S(r)} E
		_{mi}\varepsilon _k{}^{jm} x_j n^i dS &=&\; -Q(E,\mathcal{R}_k) \,.
	\end{aligned}
\end{equation}
Summary:
\begin{equation}
	\label{firstDynamic}
	\dot{Q}(E, \mathcal{T}_k) = \dot{Q}(E, \mathcal{S}) = \dot{Q}(B,
	\mathcal{T}_k) = \dot{Q}(B, \mathcal{S}) =0 \,,
\end{equation}
\begin{equation}
	\dot{Q}(E, \mathcal{R}_k) = Q(B, \mathcal{T}_k) \,,
\end{equation}
\begin{equation}
	\dot{Q}(E, \mathcal{K}_k) = Q(B, \mathcal{R}_k) \,,
\end{equation}
\begin{equation}
	\dot{Q}(B, \mathcal{R}_k) = -Q(E, \mathcal{T}_k) \,,
\end{equation}
\begin{equation}
	\label{lastDynamic}
	\dot{Q}(B, \mathcal{K}_k) = -Q(E, \mathcal{R}_k) \,.
\end{equation}

We see that for any conformal Killing field  $X$ the third time derivative of each charge mentioned above is zero:
\begin{equation}
	\frac{d^3}{dt^3} Q(E,X)=0\,.
\end{equation}
Thus, the charges defined in this way are in general quadratic polynomials of time variable.

\subsection{Solution of field equations}
\label{rozwiaznieCzesciowoNaladowane}
In \cite{1} a ``charged'' solution of field equations has been proposed. It is the simplest, non-oscillating mono-dipole solution which is singular at most on the world line at the point $r=0$.  It allows the existence of global ``potential'' (i.e., metric). We can treat it as an analogue of the Coulomb solution (only with an electric charge) in electrodynamics.\\
Let us denote by $\mathbf{p}, \mathbf{k}, \mathbf{s}$  three-dimensional vectors (which we identify with dipole functions on the sphere). Three-dimensional vector $p^k$ in Cartesian coordinates $(x^k)$ corresponds to the dipole $\mathbf{p}$, which satisfies the equation $\mathbf{p}=(p^k x_k)/r$. Similarly $\mathbf{s}=(s^k
x_k)/r$ and $\mathbf{k}= (k^l x_l)/r$.\\
The ``charged'' solution in spherical variables takes the form:
\begin{equation}
	\label{chargedfirst}
W_{BC0A} = - \frac{3}{r^2} \varepsilon_{BC}(\frac{\mathbf{s}_{,A}}{r} -
\varepsilon_A{}^D\mathbf{p}_{,D})\,,
\end{equation}
\begin{equation}
W_{AB03} = \frac{6}{r^4}\varepsilon_{AB} \mathbf{s}\,,
\end{equation}
\begin{equation}
W_{3A30} = \frac{3}{r^2}(\frac{\varepsilon_A{}^D\mathbf{s}_{,D}}{r} +
\mathbf{p}_{_A})\,,
\end{equation}
\begin{equation}
W_{3AB0} = \frac{3}{r^4} \varepsilon_{AB} \mathbf{s}\,,
\end{equation}
\begin{equation}
W_{3030} = -\frac{2}{r^3}(m + \frac{3\mathbf{k}}{r})\,,
\end{equation}
\begin{equation}
W_{0A03} = \frac{3}{r^3} \mathbf{k}_{,A}\,,
\end{equation}
\begin{equation}
W_{ABCD} = \frac{2}{r^3}(m + \frac{3\mathbf{k}}{r})(\eta_{AC} \eta_{BD} -
\eta_{AD} \eta_{BC})\,,
\end{equation}
\begin{equation}
W_{3AB3} = -W_{0AB0} = \frac{\eta_{AB}}{r^3}(m + \frac{3\mathbf{k}}{r})\,,
\end{equation}
\begin{equation}
	\label{chargedlast}
W_{BC3A} = - \frac{3}{r^3} \varepsilon_{BC}\varepsilon_A{}^D \mathbf{k}_{,D}\,,
\end{equation}
where $m$ is a monopole function, the indices $A,B,C,\dots$ correspond to the angular coordinates on the sphere, and the index $3$ to the radial coordinate. \\
It can be shown (\cite{1}) that the spin-2 field with components defined by the equations
(\ref{chargedfirst})--(\ref{chargedlast}) comes from the metric:
\begin{equation}
	h _{00} = \frac{2m}{r}+ \frac{2 \mathbf{k}}{r^2}\,,
\end{equation}
\begin{equation}
	h _{0A} = -6 \mathbf{p} _{,A} - \frac{2}{r} \varepsilon _{A}{} ^{B} \mathbf{s}
	_{,B}\,,
\end{equation}
\begin{equation}
	h _{03} = - \frac{6 \mathbf{p}}{r}\,,
\end{equation}
\begin{equation}
	h _{33} = \frac{2m}{r}+ \frac{6 \mathbf{k}}{r^2}\,,
\end{equation}
or in Cartesian coordinates $(x^k)$:
\begin{equation}
	h _{00} = \frac{2m}{r}+ \frac{2k_mx^m}{r^2}\,,
\end{equation}
\begin{equation}
	h _{0k} = -\frac{6p_k}{r}- \frac{2}{r^3} \varepsilon _{klm}s^lx^m\,,
\end{equation}
\begin{equation}
	h ^{kl} = \frac{x^k x^l}{r^2} \left( \frac{2m}{r}+ \frac{6k_mx^m}{r^3}
	\right)\,.
\end{equation}
Applying linearized Einstein equations to the above metric we obtain an  energy-momentum tensor (as a distribution located in the center of the coordinate system):
\begin{equation}
	T ^{00} = m {\boldsymbol\delta} -k^m \boldsymbol\delta _{,m}\,,
\end{equation}
\begin{equation}
	T ^{0k} = p^k \boldsymbol\delta + \frac{1}{2} \varepsilon ^{kml}s_l\boldsymbol\delta _{,m}\,,
\end{equation}
\begin{equation}
	T ^{kl}=0\,,
\end{equation}
where $\boldsymbol\delta$ is a three-dimensional Dirac delta, and $\varepsilon^{kml}$ is a
three-dimensional antisymmetric tensor ($\varepsilon^{xyz}=1$).

\subsection{A solution with additional charges}
The ``charged'' solution can be generalized by introducing additional charges.
As before, we denote: $\mathbf{q}= (q^kx_k)/r$,
$\mathbf{w} = (w^kx_k)/r$, $\mathbf{d}= (d^k x_k)/r$. We call this solution
``a fully charged solution''. It is an analogue of the Coulomb solution with an electric and magnetic charge. In contrast to the solution in the section
\ref{rozwiaznieCzesciowoNaladowane},  it does not come from a globally defined, singular (except $r=0$) metric. \\
The spin-2 field components for a fully charged solution in a spherical system (\cite{1}):
\begin{equation}
	\label{firstchargedequation}
W_{BC0A} = \varepsilon_{BC} (\frac{3}{2r} \mathbf{q}_{,A} +
\frac{3}{r^2}\varepsilon_A{}^D\mathbf{p}_{,D} - \frac{3}{r^3} \mathbf{s}_{,A})\,,
\end{equation}
\begin{equation}
W_{AB03} = \varepsilon_{AB}(\frac{3\mathbf{q}}{r^2} + \frac{2b}{r^3} +
\frac{6\mathbf{s}}{r^4})\,,
\end{equation}
\begin{equation}
W_{3A30} = -\frac{3}{2r}\varepsilon_A{}^D\mathbf{q}_{,D} + \frac{3}{r^2}
\mathbf{p}_{,A} + \frac{3}{r^3}\varepsilon_A{}^D \mathbf{s}_{,D}\,,
\end{equation}
\begin{equation}
W_{3AB0} = \varepsilon_{AB}(\frac{3\mathbf{q}}{2r^2} +
\frac{b}{r^3}+\frac{3\mathbf{s}}{r^4})\,,
\end{equation}
\begin{equation}
W_{3003} = \frac{3\mathbf{w}}{r^2} + \frac{2m}{r^3}+\frac{6\mathbf{k}}{r^4}\,,
\end{equation}
\begin{equation}
W_{A003} = \frac{3}{2r}\mathbf{w}_{,A} - \frac{3}{r^3}\mathbf{k}_{,A} -
\frac{3}{r^2}\varepsilon_A{}^C\mathbf{d}_{,C}\,,
\end{equation}
\begin{equation}
W_{ABCD} = (\frac{3\mathbf{w}}{r^2} + \frac{2m}{r^3} +
\frac{6\mathbf{k}}{r^4})\varepsilon_{AB}\varepsilon_{CD}\,,
\end{equation}
\begin{equation}
W_{3AB3}=-W_{0AB0}=\eta_{AB}(\frac{3\mathbf{w}}{2r^2}+\frac{m}{r^3}+\frac{3\mathbf{k}}{r^4})\,,
\end{equation}
\begin{equation}
	\label{lastchargedequation}
W_{3ABC}=\varepsilon_{BC}\left(\frac{3}{2r}\varepsilon_A{}^D\mathbf{w}_{,D}+\frac{3}{r^2}\mathbf{d}_{,A}
-\frac{3}{r^3}\varepsilon_A{}^D\mathbf{k}_{,D}\right)\,,
\end{equation}
where $b$ and $m$ are monopole functions.
In appendix \ref{uzupelnienie} the equations
(\ref{firstchargedequation})--(\ref{lastchargedequation}) are written in a Cartesian system. \\
The charges $\mathbf{q}$ and $\mathbf{w}$ correspond to metric tensors that do not disappear at spatial infinity.\footnote{Asymptotic of the components of the metric $ h _{\mu\nu}$ in $\frac{1}{r}$ expansion is at most a constant term, i.e. $O(1)$).} \\
It can be shown that the solution with a non-zero charge $b$ comes from the metric:
\begin{equation}
	h _{0\phi}=4b \cos\theta\,.
\end{equation}
A solution with a non-zero charge $\mathbf{d}$, where the direction $\mathbf{d}$
is parallel to the $z$ axis ($\mathbf{d}=d\cos\theta$) corresponds to a singular metric:
\begin{equation}
		h _{\theta\phi}=2rd\sin\theta\cos\theta \\
\end{equation}
or
\begin{equation}
	h _{r\theta}= 2d \left( \sin^2\theta \log \tan \frac{\theta}{2}-\cos\theta
	\right)\,.
\end{equation}
Using the definitions (\ref{linearE})--(\ref{linearB})
we calculate the electrical and magnetic parts for the full charged solution.\\
\noindent Electric part:
\begin{equation}
\begin{aligned}
			E_{ij} =&E(n_i\partial_r+y^A{}_{,i}\partial_A, n_j\partial_r +
			y^B{}_{,j}\partial_B) =&\\
			=& -\frac{\eta_{ij}}{2}\left(\frac{3\mathbf{w}}{r^2} +
			\frac{2m}{r^3} + \frac{6\mathbf{k}}{r^4}\right) - \frac{3}{r^2}n^k
			\mathbf{d}_{,l}(\varepsilon_{kj}{}^ln_i+\varepsilon_{ki}{}^ln_j)
			+&\\
			&+\frac{3}{2r^2}(n_iw_j + n_j w_i) - \frac{3}{r^4}(n_ik_j+n_jk_i) +&\\
			&- n_in_j
			\left(-\frac{3\mathbf{w}}{2r^2} - \frac{3m}{r^3} - \frac{15\mathbf{k}}{r^4}\right)\,,
\end{aligned}
\end{equation}
where $\partial_i=\frac{\partial}{\partial x_i}$  and $n_i=\frac{x_i}{r}$, or in an equivalent form:
\begin{equation}
			\begin{aligned}
			E_{ij} =&-\frac{m}{r^3}(\eta_{ij}-3n_in_j)+&\\
			&-d^l\frac{3}{r^3}n^k(\varepsilon_{kjl}n_i+\varepsilon_{kil}n_j)+& \\
			&-k^l\frac{3}{r^4}(n_l\eta_{ij}+n_i\eta_{lj}+n_j\eta_{li}-5n_in_jn_l)+& \\
			&-w^l\frac{3}{2r^2}(\eta_{ij}n_l-n_i\eta_{jl}-n_j\eta_{il}-n_in_jn_l)\,,
			\end{aligned}
\end{equation}
after writing with derivatives of the function $1/r$:
\begin{equation}
			\begin{aligned}
			E_{ij}=&\quad m\left(\frac{1}{r}\right)_{,ij}
			-k^l\left(\frac{1}{r}\right)_{,ijl}
			-[(d\times\nabla)_j\nabla_i+(d\times\nabla)_i\nabla_j)]\frac{1}{r}+ \\
			&-w^l\frac{3}{2r^2}(\eta_{ij}n_l-n_i\eta_{jl}-n_j\eta_{il}-n_in_jn_l)\,.
			\end{aligned}
\end{equation}
The magnetic part for a fully charged solution in the Cartesian system:
\begin{equation}
	\begin{aligned}
			B_{ij} =&B(n_i\partial_r+y^A{}_{,i}\partial_A, n_j\partial_r +
			y^B{}_{,j}\partial_B) =&\\
			=& -\frac{\eta_{ij}}{2}\left(\frac{3\mathbf{q}}{r^2} +
			\frac{2b}{r^3} + \frac{6\mathbf{s}}{r^4}\right) + \frac{3}{r^2}n^k
			\mathbf{p}_{,l}(\varepsilon_{kj}{}^ln_i+\varepsilon_{ki}{}^ln_j)
			+&\\
			&+\frac{3}{2r^2}(n_iq_j + n_j q_i) - \frac{3}{r^4}(n_is_j+n_js_i) +&\\
			&+ n_in_j
			\left(\frac{3\mathbf{q}}{2r^2} + \frac{3b}{r^3} + \frac{15\mathbf{s}}{r^4}\right)
	\end{aligned}
\end{equation}
or equivalently:
\begin{equation}
\begin{aligned}
			B_{ij}=&-\frac{b}{r^3}(\eta_{ij}-3n_in_j)+&\\
			&+p^l\frac{3}{r^3}(n^k(\varepsilon_{kjl}n_i+\varepsilon_{kil}n_j))+& \\
			&-s^l\frac{3}{r^4}(n_l\eta_{ij}+n_i\eta_{lj}+n_j\eta_{li}-5n_in_jn_l)+& \\
			&-q^l\frac{3}{2r^2}(\eta_{ij}n_l-n_i\eta_{jl}-n_j\eta_{il}-n_in_jn_l)\,,
\end{aligned}
\end{equation}
after writing with derivatives of the function $1/r$:
\begin{equation}
\begin{aligned}
			B_{ij}=&\quad b\left(\frac{1}{r}\right)_{,ij}
			-s^l\left(\frac{1}{r}\right)_{,ijl}
			+p^l\frac{3}{r^3}n^k(\varepsilon_{kjl}n_i+\varepsilon_{kil}n_j)+& \\
			&-q^l\frac{3}{2r^2}(\eta_{ij}n_l-n_i\eta_{jl}-n_j\eta_{il}-n_in_jn_l)\,.
\end{aligned}
\end{equation}
	Note that substituting in the expression for the electrical part: $w\rightarrow q$,
	$k\rightarrow s$, $d \rightarrow -p$,
	$m \rightarrow b$ we get a magnetic part.\\
	We can use expressions defining charges as integrals of contractions of an electrical or magnetic part with conformal Killing vectors. Calculating a fully charged solution we obtain:
	\begin{equation}
		\label{firstintegral}
	Q(E,\mathcal{S})=\int\limits_{S(r)}E_{ij}\mathcal{S}^jn^i dS = 8\pi m\,,
	\end{equation}
	\begin{equation}
	Q(E,\mathcal{T}_k)=\int\limits_{S(r)}E_{ij}\mathcal{T}^j_kn^i dS=8\pi w_k\,,
	\end{equation}
	\begin{equation}
	Q(E,\mathcal{R}_k)=\int\limits_{S(r)}E_{ij}\mathcal{R}^j_kn^i dS= -8\pi
	d_k\,,
	\end{equation}
	\begin{equation}
	Q(E,\mathcal{K}_k)=\int\limits_{S(r)}E_{ij}\mathcal{K}^j_kn^i dS= 8\pi k_k\,,
	\end{equation}
	\begin{equation}
	Q(B,\mathcal{S})=\int\limits_{S(r)}B_{ij}\mathcal{S}^jn^i dS = 8\pi b\,,
	\end{equation}
	\begin{equation}
	Q(B,\mathcal{T}_k)=\int\limits_{S(r)}B_{ij}\mathcal{T}^j_kn^i dS= 8\pi q_k\,,
	\end{equation}
	\begin{equation}
	Q(B,\mathcal{R}_k)=\int\limits_{S(r)}B_{ij}\mathcal{R}^j_kn^i dS= 8\pi p_k\,,
	\end{equation}
	\begin{equation}
		\label{lastintegral}
	Q(B,\mathcal{K}_k)=\int\limits_{S(r)}B_{ij}\mathcal{K}^j_kn^i dS= 8\pi s_k\,.
	\end{equation}
Intermediate results leading to the above equations are written in the appendix
\ref{uzupelnienie}. \\
Using dynamical equations (\ref{firstDynamic})--(\ref{lastDynamic})
we can examine dependence of charges on time variable. We receive eight charges which are time-independent, six charges which are linear in time, and six charges which are quadratic functions of time.\\
Constant in time values are:
	\begin{equation}
		m(t)=m(0), \quad w_l(t)=w_{l}(0) ,\quad b(t)=b(0) ,\quad
		q_l(t)=q_{l}(0)\,.
	\end{equation}
	Linear in time are:
	\begin{equation}
		p_l(t) = -tw_{l}(0) + p_{l}(0), \qquad d_l(t) = -tq_{l}(0) + d_{l}(0)\,.
	\end{equation}
	Quadratic in time are:
	\begin{equation}
		k_l(t) = -\frac{1}{2}t^2w_{l}(0) + tp_{l}(0) +k_{l}(0)\,,
	\end{equation}
	\begin{equation}
		s_l(t) = -\frac{1}{2}t^2q_{l}(0) +td_{l}(0) +s_{l}(0)\,,
	\end{equation}
where $m (0)$ means the value of $m$ at the initial moment $t = 0$, similarly for the remaining quantities.

%%%%%%%%%%%%%%%%%%%%%%%%%%%%%%%%%%%%%%%%%%%%%%%%%%%%%%%%%%%%%%%%%%%%%%%%%%%%%%%%
%                             Ładunki kwazilokalne                             %
%%%%%%%%%%%%%%%%%%%%%%%%%%%%%%%%%%%%%%%%%%%%%%%%%%%%%%%%%%%%%%%%%%%%%%%%%%%%%%%%

\section{Quasilocal charges in General Relativity}
\label{ladunkiKwazilokalne}

\subsection{Introduction}
In the subsection \ref{introWeyl} we have shown how one can define two symmetrical, traceless and spatial tensors of the electrical and magnetic parts of Weyl tensor. Let us assume that we have a given spatial hypersurface
$\Sigma$ equipped with a Riemannian metric $\gamma$, which we assume to be conformally flat. In the subsection \ref{introCKV} we have shown that such a hypersurface has ten linearly independent conformal Killing vectors. Now we show that the contractions of conformal Killing vectors with the electrical or magnetic part of Weyl tensor lead to objects that we can call charges of the gravitational field. Since we have ten conformal Killing vectors and two tensors, $E$ and $B$, we can define twenty
basic charges. \\
We will call a (local) charge the integral of the conformal Killing vector $X$
contracted with the electrical part $E$ or the magnetic part $B$ over a two-dimensional closed surface~$A$:
\begin{equation}
	Q(E,X):= \int_A E ^{i}{}_{j}X^j dS_i\,,
\end{equation}
\begin{equation}
	Q(B,X):= \int_A B ^{i}{}_{j}X^j dS_i\,.
\end{equation}
The charge defined in the above manner will not depend on the choice of a two-dimensional surface if the appropriate divergence is zero. If we consider two two-dimensional, oriented closed surfaces $A_1$, $A_2$ limiting the three-dimensional volume $V: \partial V = A_1 \cup A_2$, then the conformal Killing vector $X^i$ and the electrical part $E _{ij}$ fulfill the following equation:
\begin{equation}
	\label{powierzchnieS}
	\int_{A_1} \sqrt{\gamma} E ^{i}{}_{j}X^j dS_i - \int_{A_2} \sqrt{\gamma} E
	^{i}{}_{j}X^j dS_i = \int_V (\sqrt{\gamma} E ^{i}{}_{j}X^j)_{,i}dV\,.
\end{equation}
The flow calculated by any area $A_1$  will be equal to the flow through the area $A_2$ if the divergence on the right side of (\ref{powierzchnieS})
is zero. The object presented in brackets is a vector density, therefore the partial derivative can be changed into a covariant one:
\begin{equation}
	\label{divergencecalculation}
	\begin{aligned}
		(\sqrt{\gamma}E ^{i}{}_{j}X^j) _{,i} &= 	(\sqrt{\gamma}E ^{i}{}_{j}X^j) _{|i} =
	\sqrt{\gamma}E ^{i}{}_{j|i}X^j + \sqrt{\gamma} E ^{ij}X _{j|i} = \\
	&=\sqrt{\gamma} E ^{i}{}_{j|i}X^j + \sqrt{\gamma} E ^{ij}X _{(j|i)} =
   \sqrt{\gamma} E ^{i}{}_{j|i}X^j + \sqrt{\gamma} E ^{ij} \frac{\lambda}{2}g _{ij} =\\
	&=\sqrt{\gamma} E ^{i}{}_{j|i}X^j\,,
	\end{aligned}
\end{equation}
where we used the assumption that $E _{ij}$ is symmetric and traceless, and that $X^i$ is a conformal Killing vector, i.e. it satisfies the equation $X_{(i|j)}=\frac{\lambda}{2}\gamma _{ij}$.\\
The above reasoning shows that if the three-dimensional covariant divergence of the electrical part is equal to zero, the charge does not depend on the choice of
two-dimensional surface. The analogue calculation allows one to get an identical conclusion for charges defined with the help of a magnetic part. 

\noindent In the further part of this section we show how from a
three-dimensional Riemannian metric $\gamma$ and a tensor of the extrinsic curvature $K$ (initial data on the spatial hypersurface $\Sigma_0$) the electrical and magnetic parts can be reproduced. We will assume that spacetime fulfills
vacuum Einstein equations with a cosmological constant. We also show the relations of ``electromagnetic'' charges with the canonical ADM momentum and we prove formulae allowing to express the divergence of the magnetic (electrical) part by the electric (magnetic) part and the extrinsic curvature of the surface $\Sigma$.

%%%%%%%%%%%%%%%%%%%%%%%%%%%%%%%%%%%%%%%%%%%%%%%%%%%%%%%%%%%%%%%%%%%%%%%%%%%%%%%%
%                             Prozniowe bez stalej                             %
%%%%%%%%%%%%%%%%%%%%%%%%%%%%%%%%%%%%%%%%%%%%%%%%%%%%%%%%%%%%%%%%%%%%%%%%%%%%%%%%

\subsection{Vacuum equations in the (3 + 1) decomposition}
In \cite{2} the electrical and magnetic parts of Weyl tensor are expressed in terms of the metric $\gamma _{ij}$ on the surface
$\Sigma$ and
the extrinsic curvature $K _{ij}$, simultaneously assuming that spacetime fulfills the Einstein vacuum equations. Let us formulate this result in the form of a theorem and provide proof.
\begin{theorem}
\label{theoremvac}
If spacetime fulfills the Einstein vacuum equations:
\begin{equation}
	R _{\mu \nu}-\frac{1}{2}R g _{\mu \nu}=0\,,
\end{equation}
then
%w rozkładzie 3+1
the electrical and magnetic part of the Weyl tensor can be expressed as a function of the initial data $(\gamma _{ij}, K _{ij})$ specified on the three-dimensional hypersurface $\Sigma$ as follows:
\begin{equation}
	\label{lematEprozniowe}
	E _{i j} = -\overset{3}{R}{}_{i j} - K K _{i j} +
	K _{i k}K ^{k}{}_{j}\,,
\end{equation}
\begin{equation}
	\label{lematBprozniowe}
	B _{ij} = \varepsilon _{i}{}^{kl}D _{l}K _{kj}\,,
\end{equation}
where $K _{ij}$ is the extrinsic curvature tensor, $K=\gamma_{ij}K ^{ij}$, and
$\overset{3}{R}{}_{i j}$ is a three-dimensional Ricci tensor.
\end{theorem}
\noindent \textit{Proof:} \;
First, we prove the formula (\ref{lematEprozniowe}). Note that if the Einstein vacuum equations are satisfied, then the four-dimensional Ricci tensor $R _{\mu
\nu}$ is identically equal to zero, and therefore the Weyl tensor is equal to the Riemann tensor. \\
Let us use the definition of the electric part of Weyl tensor:
\begin{equation}
\label{proznioweEE}
	E _{\alpha \beta} = n^\mu n^\nu W _{\alpha \mu \nu \beta}=
	n^\mu n^\nu R _{\alpha \mu \nu \beta} =
	-n^\mu n^\nu R _{\alpha \mu  \beta\nu}\,.
\end{equation}
We can write the right hand side using a contracted Gaussian relation (\ref{contractedGauss}), and omitting the parts proportional to the four-dimensional Ricci tensor we get:
\begin{equation}
	\gamma _{\alpha \mu}n^\nu\gamma ^{\rho}{}_{\beta}n^\sigma
	\overset{4}R{}^{\mu}{}_{\nu\rho\sigma}=\overset{3}{R}{}_{\alpha\beta}+
	K K _{\alpha \beta} - K _{\alpha \mu} K ^{\mu}{}_{\beta}\,.
\end{equation}
We rewrite the left hand side with the help of a relationship between a four-dimensional metric and a three-dimensional metric, and use the symmetries of the Riemann tensor:
\begin{equation}
	\gamma _{\alpha \mu}n^\nu\gamma ^{\rho}{}_{\beta}n^\sigma
	\overset{4}{R}{}^{\mu}{}_{\nu\rho\sigma}=(g _{\alpha \mu} + n _{\alpha}
	n_\mu)(g ^{\rho}{}_{\beta}+n^\rho n_\beta)n^\sigma n^\nu
	\overset{4}{R}{}^{\mu}{}_{\nu\rho\sigma}=n^\sigma n^\nu \overset{4}{R}{}_{\alpha \nu
	\beta\sigma}\,.
\end{equation}
After converting the indices, the right hand side of the above result is proportional to the right hand side of (\ref{proznioweEE}), thus:
\begin{equation}
	E _{\alpha \beta} = -\overset{3}{R}{}_{\alpha\beta}-
	K K _{\alpha \beta} + K _{\alpha \mu} K ^{\mu}{}_{\beta}\,.
\end{equation}
Let us prove the second part of the theorem. We start again with the definition:
\begin{equation}
	B _{\alpha \beta} = W^*{}_{\alpha \mu \nu \beta}n^\nu n^\mu = \frac{1}{2}W
	_{\alpha\mu \lambda \sigma}\varepsilon ^{\lambda \sigma}{}_{\nu \beta}n^\mu
	n^\nu	=  \frac{1}{2}R
		_{\alpha\mu \lambda \sigma}\varepsilon ^{\lambda \sigma}{}_{\nu \beta}n^\mu
			n^\nu	\,.
\end{equation}
Considering the right hand side of the equality above, with the help of the Codazzi relation (\ref{codazzirelation}) we obtain:
\begin{equation}
	\gamma ^{\gamma}{}_{\rho}n^\sigma\gamma ^{\mu}{}_{\alpha}\gamma
	^{\nu}{}_{\beta}\overset{4}{R}{}^{\rho}{}_{\sigma \mu \nu}=D_\beta K
	^{\gamma}{}_{\alpha}-D_\alpha K ^{\gamma}{}_{\beta}\,.
\end{equation}
After developing the elements containing the three-dimensional metric and the reduction resulting from the symmetry of the Riemann tensor, we obtain an equivalent form:
\begin{equation}
\label{wynikzcodazziego}
	R _{\alpha \mu \lambda \sigma}n^\mu = - n^\mu n^\sigma n_\lambda R _{\alpha
	\mu \rho \sigma}-n^\mu n^\rho n_\sigma R _{\alpha \mu \lambda \rho} + D
	_{\sigma}K _{\alpha \lambda}-D_\lambda K _{\alpha \sigma}\,,
\end{equation}
thus:
\begin{equation}
	\begin{aligned}
	B _{\alpha \beta} = \frac{1}{2} \varepsilon ^{\lambda \sigma}{}_{\nu
	\beta}n^\nu&(- n^\mu n^\sigma n_\lambda R _{\alpha
	\mu \rho \sigma}-n^\mu n^\rho n_\sigma R _{\alpha \mu \lambda \rho} + D
	_{\sigma}K _{\alpha \lambda}-D_\lambda K _{\alpha \sigma} )\,.
	\end{aligned}
\end{equation}
Simplifying the contraction of the products of normal vectors with the antisymmetric tensor and omitting antisymmetrization in the elements containing the extrinsic curvature we get:
\begin{equation}
	\begin{aligned}
	B _{\alpha \beta} =  \varepsilon ^{\lambda \sigma}{}_{\nu
	\beta}n^\nu D _{\sigma}K _{\alpha \lambda} =  \varepsilon ^{\lambda \sigma \nu}{}_{
	\beta}n_\nu D _{\sigma}K _{\alpha \lambda} =-\varepsilon ^{\lambda \sigma}{}_{
	\beta}{}^\nu n_\nu D _{\sigma}K _{\alpha \lambda} \,.
	\end{aligned}
\end{equation}
We perform a contraction $n_\mu$ with an antisymmetric tensor, by definition $n_\mu$ has only a null component equal to $-N$. The remaining indices in the antisymmetric tensor
can be converted into three-dimensional, and the factor $N$
coming from the contraction allows us to change the density from four-dimensional to three-dimensional:
\begin{equation}
	B _{ab} = \varepsilon ^{ls}{}_{b}D_s K _{al}\,.
\end{equation}
Initially, the magnetic part was defined as a symmetrical tensor, so we can write:
\begin{equation}
	B _{ab} = \varepsilon ^{ls}{}_{a}D_s K _{bl} = \varepsilon _a {}^{ls} D_s K
	_{bl}\,,
\end{equation}
which ends the proof of the theorem.\qed

%%%%%%%%%%%%%%%%%%%%%%%%%%%%%%%%%%%%%%%%%%%%%%%%%%%%%%%%%%%%%%%%%%%%%%%%%%%%%%%%
%                       Prozniowe ze stala kosmologiczna                       %
%%%%%%%%%%%%%%%%%%%%%%%%%%%%%%%%%%%%%%%%%%%%%%%%%%%%%%%%%%%%%%%%%%%%%%%%%%%%%%%%

\subsection{Vacuum equations with a cosmological constant}
Analyzing the proof of the theorem \ref{theoremvac}, we can notice that a slight modification of the formulae defining the electrical and magnetic part can lead to the generalization of theorem on vacuum spacetimes with a non-zero cosmological constant.

\begin{theorem}
	\label{theoremEBlambda}
If spacetime fulfills the Einstein vacuum equations with a
cosmological constant:
\begin{equation}\label{tw2Einsteinvaceq}
	R _{\mu \nu}-\frac{1}{2}R g _{\mu \nu}+\Lambda g _{\mu\nu}=0\,,
\end{equation}
the electrical and magnetic part of the Weyl tensor is expressed by the initial data
$(\gamma _{ij}, K _{ij})$ on the three-dimensional spatial hypersurface
$\Sigma$ as follows:
\begin{equation}
	\label{lematElambda}
	E _{i j} = -\overset{3}{R}{}_{i j} - K K _{i j} +
	K _{i k}K ^{k}{}_{j}+\frac{2}{3}\Lambda \gamma_{ij}\,,
\end{equation}
\begin{equation}
	\label{lematBlambda}
	B _{ij} = \varepsilon ^{ls}{}_{j}D_s K _{il}\,,
\end{equation}
where $\overset{3}{R}{}_{ij}$ is a Ricci tensor of a three-dimensional metric, $K
_{ij}$ is a tensor of extrinsic curvature, and  $K=K ^{i}{}_{i}$ is a trace of extrinsic curvature.
\end{theorem}
\noindent\textit{Proof:}\;
We carry out the proof for the electrical part first. Riemann tensor can be splitted into Weyl tensor and traces of Riemann tensor.
In four dimensions, the Weyl tensor takes the following form:
\begin{equation}
\label{weyl}
	W _{\alpha \beta \mu \nu} = R _{\alpha \beta \mu \nu}-(g _{\alpha [ \mu}R
	_{\nu]\beta}- g _{\beta[\mu}R _{\nu]\alpha}) + \frac{1}{3}R g _{\alpha [\mu}g
	_{\nu]\beta}\,.
\end{equation}
The definition of the electrical part:
\begin{equation}
\label{theoremcosmo1}
	\begin{aligned}
		-E _{\alpha \beta}=&-n^\mu n^\nu W _{\alpha \mu \nu \beta} = n^\mu n^\nu W
		_{\alpha \mu \beta \nu} =& \\
		=& n^\mu n^\nu R _{\alpha \mu \beta \nu} - \frac{1}{2}\left(g _{\alpha \beta}R
		_{\mu \nu}n^\mu n^\nu - g _{\alpha \nu}R _{\beta \mu}n^\beta n^\mu - g
		_{\mu \beta}R _{\alpha \nu}n^\alpha n^\nu + g _{\mu \nu}R _{\alpha
		\beta}n^\mu n^\nu\right) +&\\
		+& \frac{1}{6}R g _{\alpha \beta}g _{\mu \nu}n^\mu n^\nu
		- \frac{1}{6}R g _{\alpha \nu}g _{\beta \mu}n^\mu n^\nu =& \\
		=& n^\mu n^\nu R _{\alpha \mu \beta \nu} - \frac{1}{2}g _{\alpha \beta}R
		_{\mu \nu}n^\mu n^\nu + \frac{1}{2}R _{\beta \mu}n^\mu n _{\alpha}+
		\frac{1}{2}R _{\alpha \nu} n _{\beta}n^\nu + \frac{1}{2}R _{\alpha
		\beta}+& \\
		-& \frac{1}{6}R g _{\alpha \beta}- \frac{1}{6}R n _{\alpha}n _{\beta}\,.
	\end{aligned}
\end{equation}
%Jeżeli teraz wykorzystamy zwężoną relację Gaussa (\ref{contractedGauss}):
Using the contracted Gaussian relation (\ref{contractedGauss}) we obtain:
\begin{equation}
\label{theoremcosmo2}
	\gamma ^{\mu}{}_{\alpha}\gamma ^{\nu}{}_{\beta}\overset{4}{R}{}_{\mu
	\nu}+ \gamma _{\alpha \mu}n^\nu\gamma ^{\rho}{}_{\beta}n^\sigma
	\overset{4}R{}^{\mu}{}_{\nu\rho\sigma}=\overset{3}{R}{}_{\alpha\beta}+
	K K _{\alpha \beta} - K _{\alpha \mu} K ^{\mu}{}_{\beta}\,.
\end{equation}
For the second term on the left hand side of equality, we use the relationship between a three-dimensional and four-dimensional metric:
\begin{equation}
\label{theoremcosmo3}
	\gamma _{\alpha \mu}\gamma ^{\rho}{}_{\beta}n^\sigma n^\nu
	\overset{4}{R}{}^{\mu}{}_{\nu\rho\sigma}=(g _{\alpha \mu} + n _{\alpha}
	n_\mu)(g ^{\rho}{}_{\beta}+n^\rho n_\beta)n^\sigma n^\nu
	\overset{4}{R}{}^{\mu}{}_{\nu\rho\sigma}=n^\sigma n^\nu \overset{4}{R}{}_{\alpha \nu
	\beta\sigma}
\end{equation}
and for Ricci tensor:
\begin{equation}
\label{theoremcosmo4}
	\gamma ^{\mu}{}_{\alpha}\gamma ^{\nu}{}_{\beta}\overset{4}{R}{}_{\mu\nu}=(g
	^{\mu}{}_{\alpha}+n^\mu n_\alpha)(g ^{\nu}{}_{\beta}+n^\nu
	n_\beta)\overset{4}{R}{} _{\mu\nu}\,.
\end{equation}
Let us also use the Einstein equations:
\begin{equation}
\label{theoremcosmo5}
	R _{\mu\nu}-\frac{1}{2}Rg _{\mu\nu}+\Lambda g _{\mu\nu}=0\,.
\end{equation}
Contracting the Einstein equations (\ref{tw2Einsteinvaceq}) gives:
\begin{equation}
\label{theoremcosmo6}
	R - 2R + 4 \Lambda =0 \quad \Rightarrow \quad R=4\Lambda\,.
\end{equation}
Equations (\ref{theoremcosmo2})--(\ref{theoremcosmo6}) allow us to express the right hand side of (\ref{theoremcosmo1}) through the right hand side of (\ref{theoremcosmo2}) plus additional terms that can be converted from Einstein equations to expression proportional to the product of the cosmological constant and three-dimensional metric. After completing all the operations, we get:
\begin{equation}
	E _{\alpha \beta} = -\overset{3}{R}{}_{\alpha \beta} - K K _{\alpha \beta} +
	K _{\alpha \mu}K ^{\mu}{}_{\beta}+\frac{2}{3}\Lambda \gamma_{\alpha
	\beta}\,.
\end{equation}
All objects appearing in the above equation are well-defined tensors on the spatial surface $\Sigma$, so we can write:
\begin{equation}
	E _{i j} = -\overset{3}{R}{}_{i j} - K K _{i j} +
	K _{i k}K ^{k}{}_{j}+\frac{2}{3}\Lambda \gamma_{i
	j}\,.
\end{equation}

For the magnetic part $ B $, by definition we have:
\begin{equation}
	B _{\alpha \beta} = W^*{}_{\alpha \mu \nu \beta}n^\nu n^\mu = \frac{1}{2}W
	_{\alpha\mu \lambda \sigma}\varepsilon ^{\lambda \sigma}{}_{\nu \beta}n^\mu
	n^\nu	\,.
\end{equation}
The equivalent form of the Codazzi relation (\ref{wynikzcodazziego})
and the definition of Weyl tensor (\ref{weyl}) yield:
\begin{equation}
	\begin{aligned}
	B _{\alpha \beta} = \frac{1}{2} \varepsilon ^{\lambda \sigma}{}_{\nu
	\beta}n^\nu&[- n^\mu n^\sigma n_\lambda R _{\alpha
	\mu \rho \sigma}-n^\mu n^\rho n_\sigma R _{\alpha \mu \lambda \rho} + D
	_{\sigma}K _{\alpha \lambda}-D_\lambda K _{\alpha \sigma} + \\
	&-(n^\mu g _{\alpha [\lambda}R _{\sigma ]\mu}-n^\mu g _{\mu [\lambda}R
	_{\sigma]\alpha}) + \frac{1}{3}R g _{\alpha[\lambda}g _{\sigma ]
	\mu}n^\mu]\,.
	\end{aligned}
\end{equation}
We can omit antisymmetry and elements that are symmetrical products of normal vectors contracted with an antisymmetric tensor:
\begin{equation}
\label{dowodB1}
	\begin{aligned}
	B _{\alpha \beta} = \frac{1}{2} \varepsilon ^{\lambda \sigma}{}_{\nu
	\beta}n^\nu\left( 2D _{\sigma}K _{\alpha \lambda}
	-n^\mu g _{\alpha \lambda}R _{\sigma \mu}
	\right)\,.
	\end{aligned}
\end{equation}
Let us use the contracted Codazzi relationship (\ref{contractedcodazzi}):
\begin{equation}
	\begin{aligned}
	&D _{\alpha}K - D _{\mu}K ^{\mu}{}_{\alpha}=\gamma ^{\mu}{}_{\alpha}n^\nu R
	_{\mu \nu}= n^\nu R _{\alpha \nu}+n^\mu n^\nu n_\alpha R _{\mu\nu} \quad
	\Rightarrow \\
	\Rightarrow \quad&
	n^\mu R _{\sigma \mu} = D _{\sigma}K - D _{\rho}K ^{\rho}{}_{\sigma}-n^\rho
	n^\delta n_\sigma R _{\rho\delta}\,.
	\end{aligned}
\end{equation}
The above result is used in (\ref{dowodB1}), omitting the element containing the products of normal vectors, which is reduced due to the contraction with the antisymmetric tensor:
\begin{equation}
	\begin{aligned}
	B _{\alpha \beta} = \frac{1}{2} \varepsilon ^{\lambda \sigma}{}_{\nu
	\beta}n^\nu\left( 2D _{\sigma}K _{\alpha \lambda}
	-g _{\alpha \lambda} D_\sigma K + g _{\alpha \lambda} D_\rho K
	^{\rho}{}_{\sigma}
	\right)\,.
	\end{aligned}
\end{equation}
We perform contraction of the antisymmetric tensor with the normal vector:
\begin{equation}
\label{wynikBcosmo}
	B _{ab} = \varepsilon ^{ls}{}_{b}D_s K _{al}+ \varepsilon ^{s}{}_{ba}(D _{k} K
	^{k}{}_{s}-D_sK)\,.
\end{equation}
The terms appearing in brackets on the right hand side are zero, using the vacuum vector constraint (with cosmological constant) we obtain:
\begin{equation}
	B _{ab} = \varepsilon ^{ls}{}_{b}D_s K _{al}\,.
\end{equation}
\qed

\noindent Note that if we consider equations with a cosmological constant and matter then:
\begin{equation}
	B _{ab}= B _{(ab)}= \varepsilon ^{ls}{}_{(b}K _{a)l|s}\,.
\end{equation}
The above equation is valid because  $B$ was defined as a symmetric tensor, so adding symmetrization in the indices  $a$, $b$ to the equation (\ref{wynikBcosmo}) reduces the proportional term to the antisymmetric tensor $\varepsilon ^{s}{}_{ba}$.

%%%%%%%%%%%%%%%%%%%%%%%%%%%%%%%%%%%%%%%%%%%%%%%%%%%%%%%%%%%%%%%%%%%%%%%%%%%%%%%%
%                       Dywergencja czesci elektrycznej                        %
%%%%%%%%%%%%%%%%%%%%%%%%%%%%%%%%%%%%%%%%%%%%%%%%%%%%%%%%%%%%%%%%%%%%%%%%%%%%%%%%

\subsection{Divergence of electrical and magnetic parts}

In the introduction, we have shown that vanishing of the three-dimensional, covariant divergence of the electrical part is a sufficient condition that the values of the respective charges are independent of the choice of the two-dimensional integration surface. \\
From the results of theorem \ref{theoremEBlambda} we can
present the divergence of electrical (magnetic) part of Weyl tensor only by magnetic (electrical) part of Weyl tensor and extrinsic curvature.
\begin{theorem}
	\label{divergenceTheorem}
If spacetime fulfills the Einstein vacuum equations with a cosmological constant, the three-dimensional covariant divergence of the electrical part $E$
and the magnetic $B$ of Weyl tensor is expressed as follows:
\begin{equation}
	E ^{i}{}_{j|i} = (K \wedge B)_{j}\,,
\end{equation}
\begin{equation}
	B ^{i}{}_{j|i} = -(K \wedge E) _{j}\,,
\end{equation}
where $\wedge$ is an operation defined for two symmetric tensors $A$ and
$B$ as
\begin{equation}
	(A \wedge B)_a :=
	\varepsilon _{a}{} ^{bc}A _{b}{}^dB _{dc}\,.
\end{equation}
\end{theorem}
\noindent
It can be especially useful in situations where we want to analyze the ``pure'' electrical (with zero magnetic part) or ``pure'' magnetic (with zero electrical part) spacetimes. Note that from the above formulation it follows that if we consider spaces with zero magnetic part, the divergence of the electrical part automatically disappears, and therefore all ``electric'' charges are independent of the choice of the integration surface. Analogously for space with zero electrical part and ``magnetic'' charges.\\
\textit{Proof of the theorem \ref{divergenceTheorem}}
Using the results of the theorem \ref{theoremEBlambda} we can express the electrical part by the three-dimensional Ricci tensor, the extrinsic curvature, the metric and the cosmological constant:
\begin{equation}
\label{divErach}
\begin{aligned}
	E ^{i}{}_{j|i}=&(-R ^{i}{}_j - KK ^{i}{}_j + K ^{i}{}_{k}K ^{k}{}_j +\frac{2}{3}\Lambda
	\gamma ^{i}{}_j) _{|i} = \\
	=&-\frac{1}{2}(K ^{kl}K _{kl}-K^2+2\Lambda)_{|j} +(- KK ^{i}{}_j + K
	^{i}{}_{k}K ^{k}{}_j) _{|i} = \\
	=& -\frac{1}{2}(K ^{kl}K _{kl}) _{|j} + K ^{i}{}_{k}K
	^{k}{}_{j|i} =
	 -K ^{kl}K _{kl} {}_{|j} + K ^{i}{}_{k}K
	^{k}{}_{j|i}=\\
	=&\quad K ^{i}{}_{k} ( K ^{k}{}_{j|i}- K ^{k}{}_{i|j}) = K ^{ik}(K
	_{kj|i}- K _{ki|j})\,.
\end{aligned}
\end{equation}
We have used the Bianchi identity and the scalar constraint equation:
\begin{equation}
	R ^{i}{}_{j|i} = \frac{1}{2}R _{|j} = \frac{1}{2} \left( K ^{ij}K _{ij} - K^2 + 2
	\Lambda \right) _{|j}\,.
\end{equation}
In the third line of (\ref{divErach}), we have simplified the expression by changing the derivative of the trace of the extrinsic curvature by the divergence of the extrinsic curvature (we assume that the vector constraint is satisfied):
\begin{equation}
	K _{|i} = K ^{j}{}_{i|j}\,.
\end{equation}
The result obtained in the calculation (\ref{divErach}) can be equivalently written as:
\begin{equation}
\label{divEprodukt}
	E ^{i}{}_{j|i} = (K \wedge B)_{j}\,.
\end{equation}
The correctness of the formula (\ref{divEprodukt}) can be checked by a direct calculation using the definition of the operation $\wedge$ and the identity allowing to write the contraction of antisymmetric symbols as a difference of products of the Kronecker delta. 

%%%%%%%%%%%%%%%%%%%%%%%%%%%%%%%%%%%%%%%%%%%%%%%%%%%%%%%%%%%%%%%%%%%%%%%%%%%%%%%%
%                        Dywergencja czesci magetycznej                        %
%%%%%%%%%%%%%%%%%%%%%%%%%%%%%%%%%%%%%%%%%%%%%%%%%%%%%%%%%%%%%%%%%%%%%%%%%%%%%%%%

\noindent Proof of the analogical relation for the magnetic part of Weyl tensor requires ``commutator'' for the second covariant derivatives. For any tensor $X
^{a}{}_{b}$ the following formula holds:
\begin{equation}
	X ^{a}{}_{b;cd}-X ^{a}{}_{b;dc}= X ^{e}{}_{b}R ^{a}{}_{edc} - X ^{a}{}_{e} R
	^{e}{}_{bdc}\,.
\end{equation}
We apply it to the magnetic part of Weyl tensor (for the form resulting from the theorem
(\ref{theoremEBlambda})); in this case, $R$ stands for the three-dimensional Riemann tensor on the spatial hypersurface $\Sigma$.
\begin{equation}
	B ^{ij}{}_{|i} = \varepsilon ^{ilk}K ^{j}{}_{l|ki} = \frac{1}{2}\varepsilon
	^{ilk}(K ^{m}{}_{l} R ^{j}{}_{mik} - K ^{j}{}_{m} R ^{m}{}_{lik})\,.
\end{equation}
The second term vanishes because the antisymmetry of the Riemann tensor in the last three indices is equal to zero, so we have:
\begin{equation}
	\label{kowdiwB}
	B ^{ij}{}_{|i}=  \frac{1}{2}\varepsilon
	^{ilk}K ^{m}{}_{l} R ^{j}{}_{mik}\,.
\end{equation}
Riemann tensor of three-dimensional space is entirely expressed by Ricci tensor and scalar of curvature:
\begin{equation}
	R _{jmik} = 2( \gamma _{j[i} R _{k]m}-\gamma _{m[i}R _{k]j}) -R \gamma
	_{j[i}\gamma _{k]m}\,,
\end{equation}
hence the following:
\begin{equation}
	\begin{aligned}
	B ^{ij}{}_{|i} &= \varepsilon ^{ilk}K ^{m}{}_{l}(\gamma ^{j}{}_{i}R _{km} - \gamma _{mi} R
	^{j}{}_{k}) - \frac{1}{2}\varepsilon ^{ilk}K ^{m}{}_{l}R \gamma ^{j}{}_{i}\gamma _{km} &=& \\
	&= \varepsilon ^{jlk}K ^{m}{}_{l} R _{km}-\varepsilon ^{ilk}K _{il} R ^{j}{}_{k} -
	\frac{1}{2} \varepsilon ^{jlk}K _{kl}R &=&\quad \varepsilon ^{jlk}K ^{m}{}_{l} R
		_{km}\,.
	\end{aligned}
\end{equation}
We now show that the divergence of the magnetic part of Weyl tensor can be written as the ``vector product'' of the electrical part and the extrinsic curvature tensor:
\begin{equation}
	B ^{i}{}_{j|i} = -(K \wedge E) _{j}\,.
\end{equation}
\begin{equation}
	\begin{aligned}
	-(K \wedge E) _{j} &= - \varepsilon _j{}^{bc} K ^{d}{}_{b} E _{dc}
	= - \varepsilon_j{}^{bc} K ^{d}{}_{b} \left( -R _{dc}-KK _{dc}+ K _{dk} K
	^{k}{}_{c} + \frac{2}{3}\Lambda \gamma _{dc} \right) = \\
	&= \varepsilon_j {}^{bc} K ^{d}{}_{b}R _{dc} = \varepsilon_j {}^{lk}K ^{m}{}_{l} R
	_{km} = B ^{i}{}_{j|i}\,.
	\end{aligned}
\end{equation}
\qed

\subsection{Linear and angular momentum}
In this subsection, we  assume that the spatial hypersurface $\Sigma$
is equipped with a flat three-dimensional metric $\gamma$. We formulate two theorems,
which enable us to give a correspondence between the expression describing the linear momentum with the ``magnetic'' charge (theorem \ref{theoremmomentum}) and the definition of the angular momentum with the integral of contraction of the magnetic part with the generator of proper conformal transformations (theorem \ref{theoremangularmomentum}).\\
%The expression "ADM moment of momentum" has been deliberately placed in quotes.
%In general, those objects do not transform like a vector. The problem can be partially %solved by limiting the initial data to $\Sigma$ with a sufficiently strong asymptotics %in spatial infinity (\cite{4}, \cite{5}).\\
%We omit a detailed discussion of this phenomenon. Our goal is to demonstrate the %equality of the relevant integrals.
\begin{theorem}
	\label{theoremmomentum}
	Let $\Sigma$ be a flat, three-dimensional spatial hypersurface immersed in the spacetime which satisfies Einstein vacuum equations.
Assuming that the three-dimensional covariant divergence of the magnetic part disappears:
	\begin{equation}
		\label{zalozenieodywergencejiliniowy}
		B ^{i}{}_{j|i}=0\,,
	\end{equation}
	then for any two-dimensional, closed surface $A$ immersed in
	$\Sigma$ holds:
	\begin{equation}
		\label{tezaLiniowy}
		\int_A B^{i}{}_{j} \mathcal{R}^{j}_{\mathbf k}dS_i =
		\int_A P^{i}{}_{j} \mathcal{T}^{j}_{\mathbf k}dS_i\,,
	\end{equation}
	where $P^{i}{}_{j}$ is (canonical) ADM momentum, $\mathcal{R}_{\mathbf k}$ is a rotation generator around the axis
	$k$, and $\mathcal{T}_{\mathbf k}$ is a translation generator along the axis $k$.
\end{theorem}
\noindent\textit{Proof:}\;
%TODO napisac ze mozna przejsc na sfere
The surface $\Sigma$ is assumed to be flat, so appropriate conformal Killing fields exist. \\
First we show that an integral over any surface $A$ can be converted into an integral over a two-dimensional sphere.\\
The ADM momentum is expressed by the extrinsic curvature: %\marginpar{\small poprawic znak pedu ADM}
\begin{equation}
	P ^{i}{}_{j} = -K ^{i}{}_{j} + K \gamma ^{i}{}_{j}\,.
\end{equation}
Einstein vacuum equations are satisfied, so in particular vacuum vector constraint:
\begin{equation}
	\label{lemmaconst}
	K ^{i}{}_{j|i}-K _{|j}=0 \quad \Rightarrow \quad P ^{i}{}_{j|i}=0	\,.
\end{equation}
The divergence in the integral in the equation (\ref{tezaLiniowy}) can be reformulated analogically to the equation (\ref{divergencecalculation}). Assuming
(\ref{zalozenieodywergencejiliniowy}), we obtain that the divergence of the term containing the magnetic part is zero. For the integral on the right hand side of the equation (\ref{tezaLiniowy}), we have:
\begin{equation}
	(\sqrt{\gamma} P ^{i}{}_{j} \mathcal{T}^{j}_{\mathbf k}) _{,i}=
	(\sqrt{\gamma} P ^{i}{}_{j} \mathcal{T}^{j}_{\mathbf k}) _{|i}=
	\sqrt{\gamma} P ^{i}{}_{j|i}\mathcal{T}^j_{\mathbf k} + \sqrt{\gamma}P ^{(ij)}  (\mathcal{T}_{\mathbf k})
	{}_{(j|i)}=0\,,
\end{equation}
where we used the fact that the divergence calculated with the help of
partial derivative is equal to covariant divergence (because the object being differentiated is the vector density). After applying the Leibniz rule, the first term in the above equation vanishes from vector constraint (\ref{lemmaconst}), and in the second we can add symmetrization and use the Killing equation for translation generator  $\mathcal{T}_{\mathbf k}$. \\
We have shown that the divergences of both integrands in equation (\ref{tezaLiniowy}) are zero, using the Stokes theorem, the volume member has no contribution, which
proves formulae (\ref{tezaLiniowy}). Replacing the integral on
any surface $A$ by an integral on a two-dimensional sphere $S(r)$ is justified.\\
Let us transform the left side of the thesis:
\begin{equation}
	\int_{S(r)} B ^{i}{}_{j} \mathcal{R} ^j_{\mathbf k} dS_i = \int_{S(r)} \lambda B
	^{r}{}_{\phi}\,,
\end{equation}
where $\lambda = r^2\sin\theta$.
Using the expression describing the magnetic part by the initial data on $\Sigma$:
\begin{equation}
	B ^{r}{}_{\phi}= \varepsilon ^{rij} K _{\phi i |j}\,.
\end{equation}
In the adopted convention for the antisymmetric tensor on the sphere: $r^2\sin\theta \varepsilon
^{\theta\phi}=1$, so we have:
\begin{equation}
	B ^{r}{}_{\phi} = \varepsilon ^{rAB} K _{\phi A | B} r^2 \varepsilon ^{\phi
	C}(\cos\theta) _{,C} = \varepsilon ^{AB}K _{DA|B}r^2 \varepsilon ^{DC} (\cos\theta)
	_{,C}\,.
\end{equation}
In the above we have exchanged three-dimensional indices for angular coordinates on the sphere. A three-dimensional covariant derivative can be decomposed into:
\begin{equation}
	K _{DA|B} = K _{DA,B} - \Gamma ^{m}{}_{DB}K _{mA} - \Gamma ^{m}{}_{AB}K
	_{Dm}= K _{DA||B} + \frac{1}{r}\eta _{AB}K _{Dr}+\frac{1}{r}\eta _{DB}K
	_{rA}\,,
\end{equation}
where we used the fact that $\Gamma ^{r}{}_{AB} = - \frac{1}{r}\eta _{AB}$,
true for a covariant derivative of a flat three-dimensional space, and $\eta
_{AB}$ is a metric induced on a sphere with the radius $r$, so:
\begin{equation}
	B ^{r}{}_{\phi} = \varepsilon ^{AB}K _{DA||B}r^2 \varepsilon ^{DC} (\cos\theta)
	_{,C}+r \varepsilon ^{A}{}_{D}K _{rA} \varepsilon ^{DC} (\cos\theta) _{,C}\,.
\end{equation}
For a two-dimensional Levi-Civita tensor, the following identity is true:
%\newline
$\varepsilon ^{A}{}_{D}\varepsilon ^{DC}= -\eta ^{AC}$. Further transformation of the magnetic part:
\begin{equation}
	\begin{aligned}
		B ^{r}{}_{\phi} &= \varepsilon ^{AB}K _{DA||B}r^2 \varepsilon ^{DC} (\cos\theta)
	_{,C}-r K _{r}{}^C  (\cos\theta) _{,C} = \\
	&=\varepsilon ^{AB}K _{DA||B}r^2 \varepsilon ^{DC} (\cos\theta)
	_{,C}+r K _{r}{}^\theta  \sin\theta = \\
	&=\varepsilon ^{AB}K _{DA||B}r^2 \varepsilon ^{DC} (\cos\theta)
	_{,C}+ \frac{1}{r} K _{r\theta}  \sin\theta\,.
	\end{aligned}
\end{equation}
The above result is integrated on the sphere of the radius $r$:
\begin{equation}
	\label{linearmomentumlemma}
	\begin{aligned}
		\int_{S(r)} \lambda B ^{r}{}_{\phi} &=
	\int_{S(r)}\lambda\varepsilon ^{AB}K _{DA||B}r^2 \varepsilon ^{DC} (\cos\theta)
	_{,C}+ \int_{S(r)}\lambda \frac{1}{r} K _{r\theta}  \sin\theta = \\
	=&-\int_{S(r)}\lambda\varepsilon ^{AB}K _{DA}r^2 \varepsilon ^{DC} (\cos\theta)
	_{||CB}+ \int_{S(r)}\lambda \frac{1}{r} K _{r\theta}  \sin\theta = \\
	=&\int_{S(r)}\lambda\varepsilon ^{AB}K _{DA} \varepsilon ^{DC}
	\eta _{CB}\cos\theta+ \int_{S(r)}\lambda \frac{1}{r} K _{r\theta}  \sin\theta = \\
	=&\int_{S(r)}\lambda K _{DA}  \eta^{DA}
	\cos\theta+ \int_{S(r)}\lambda \frac{1}{r} K _{r\theta}  \sin\theta \,,
	\end{aligned}
\end{equation}
where we used integration by parts and the identity $r^2(\cos\theta)
_{||CB} = -\eta _{CB}\cos\theta$.\\
%Let's write down the second integral in the formulation of the thesis, we will need an %expression for
The field $\mathcal{T}_z$ in the spherical coordinates:
\begin{equation}
	\mathcal{T}_z = \partial_z = \cos\theta \partial_r -
	\frac{\sin\theta}{r}\partial_\theta\,.
\end{equation}
Therefore, the right hand side of the thesis (\ref{tezaLiniowy}) takes the form:
\begin{equation}
	\begin{aligned}
		-\int_{S(r)} P ^{i}{}_{j} \mathcal{T}_z^j dS_i &= \int_{S(r)} \left( K
		^{i}{}_{j} - \delta ^{i}{}_{j} K \right) \mathcal{T}_z^j dS_i &=& \\
	&= \int_{S(r)} \lambda (K ^{r}{}_{r}-K)\cos\theta -\lambda
		\frac{1}{r}\sin\theta K ^{r}{}_{\theta} &=& \\
	&= \int_{S(r)}  -K ^{AB}\eta _{AB}\lambda\cos\theta -\lambda
		\frac{1}{r}\sin\theta K {}_{r\theta} &=&\quad - \int_{S(r)} \lambda B
		^{r}{}_{\phi}\,,
	\end{aligned}
\end{equation}
where the last equality comes from the comparison with the right hand side of
(\ref{linearmomentumlemma}).
\qed
\begin{theorem}
	\label{theoremangularmomentum}
	Let $\Sigma$ be a flat, three-dimensional spatial hypersurface immersed in the spacetime which satisfies Einstein vacuum equations. Assuming that the following charges disappear:
	\begin{equation}
		\label{zerowezwezenia}
		Q(B,\mathcal{T}_x) = Q(B, \mathcal{T}_y) = Q(B, \mathcal{T}_z)=0
	\end{equation}
	and the three-dimensional covariant divergence of the magnetic part disappears:
	\begin{equation}
		\label{zalozenieodywergenceji}
		B ^{i}{}_{j|i}=0\,,
	\end{equation}
	then for any two-dimensional, closed surface $A$ immersed in
	$\Sigma$ holds:
	\begin{equation}
		\label{teza}
		\int_A B ^{i}{}_{j} \mathcal{K} ^{j}_{\mathbf k}dS_i =
		\int_A P ^{i}{}_{j} \mathcal{R} ^{j}_{\mathbf k}dS_i\,,
	\end{equation}
	where $P^{i}{}_{j}$ is the ADM momentum, $\mathcal{K}_{\mathbf k}$ is the generator of proper conformal transformations in the direction of $k$, and
	$\mathcal{R}_{\mathbf k}$ is a rotation generator around the axis $k$.
\end{theorem}
\noindent\textit{Proof:}\;
Analogically to the proof of theorem \ref{theoremmomentum}, it can be shown that the divergences of integrands in the thesis (\ref{teza})
disappear, that justifies the proof for integrals on
two-dimensional spheres. \\
By assumption (\ref{zerowezwezenia}), we can deduce that the integral from the contraction of the magnetic part with any vector with constant coefficients in the
Cartesian system is zero (because $\mathcal{T}_{\mathbf k} = \partial_k$). Using
this observation, we can write:
\begin{equation}
	\label{stalywektor}
	\int_{S(r)} B ^{i}{}_{j}( \mathcal{K}^j_{\mathbf k} + A^j)dS_i = 	
	\int_{S(r)} B ^{i}{}_{j} \mathcal{K}^j_{\mathbf k}dS_i  	\,,
\end{equation}
where $A^j$ are the coordinates of a constant vector in the Cartesian system. Any
such a vector can be written in a spherical system in the following way:
\begin{equation}
	A^i \partial_i = \frac{A_ix^i}{r}\partial_r + r \left(
	\frac{A_i x^i}{r}
	\right)^{,B}\partial_B\,.
\end{equation}
We denote $u:= (A_ix^i)/r$, then:
\begin{equation}
	A^i \partial_i = u\partial_r + r\cdot (u)^{,B}\partial_B\,.
\end{equation}
Now we choose $u$ in such way to simplify the calculation of the integral (\ref{stalywektor}) as much as possible. Let us write the expressions for generators of the conformal transformations in the spherical system:
\begin{equation}
	\mathcal{K}_x = \frac{1}{2}r^2\cos\phi\sin\theta\partial_r -
	\frac{1}{2}r\cos\phi\cos\theta\partial_\theta+ \frac{1}{2}r
	\frac{\sin\phi}{\sin\theta}\partial_\phi\,,
\end{equation}
\begin{equation}
	\mathcal{K}_y = \frac{1}{2}r^2\sin\phi\sin\theta\partial_r -
	\frac{1}{2}r\sin\phi\cos\theta\partial_\theta - \frac{1}{2}r
	\frac{\cos\phi}{\sin\theta}\partial_\phi\,,
\end{equation}
\begin{equation}
	\mathcal{K}_z =
	\frac{1}{2}r^2\cos\theta\partial_r+\frac{1}{2}r\sin\theta\partial_\theta\,,
\end{equation}
Let us choose the function $u$:\\
For $\mathcal{K}_x$ we choose $u_x=\frac{1}{2}r^2\cos\phi\sin\theta$.\\
For $\mathcal{K}_y$ we choose $u_y=\frac{1}{2}r^2\sin\phi\sin\theta$.\\
For $\mathcal{K}_z$ we choose $u_z=\frac{1}{2}r^2\cos\theta$.\\
We will denote new ``corrected'' vectors by $\bar{\mathcal{K}}_{\mathbf k} :=
\mathcal{K}_{\mathbf k} + (A_k)^i \partial_i$, and we obtain:
\begin{equation}
	\bar{\mathcal{K}}_x = r^2\cos\phi\sin\theta\partial_r = r^2 \frac{\partial
	r}{\partial x}\partial_r\,,
\end{equation}
\begin{equation}
	\bar{\mathcal{K}}_y = r^2\sin\phi\sin\theta\partial_r = r^2 \frac{\partial
	r}{\partial y}\partial_r\,,
\end{equation}
\begin{equation}
	\bar{\mathcal{K}}_z = r^2\cos\theta\partial_r= r^2 \frac{\partial
	r}{\partial z}\partial_r\,.
\end{equation}
The vectors $\mathcal{K}_{\mathbf k}$ now have only the component in the radial direction.\\
Without loss of generality let us assume that we show equality (\ref{teza}) for $k$
equal to $z$ (we can always choose such coordinate system that the $z$ axis is turned in the direction which is invariant under the rotation generator $\mathcal{R}_{\mathbf k}$).\\
In the spherical system, the metric on $\Sigma$
takes the form:
\begin{equation}
	\gamma _{ij} =  dr^2 + r^2 d\theta^2 + r^2\sin^2\theta d\phi^2\,,
\end{equation}
and the metric induced on the spheres:
\begin{equation}
	\eta_{AB} =  r^2d\theta^2 + r^2\sin^2\theta d\phi^2 \,.
\end{equation}
Note that the rotation generator field around the $z$ axis such that $\mathcal{R}_z=\partial_\phi$ can be written as:
\begin{equation}
	\partial_\phi =  r^2 \varepsilon ^{AB} \left( \cos\theta
	\right)_{,B}\partial_A\,,
\end{equation}
where $\varepsilon ^{AB}$ is an antisymmetric tensor on the sphere, by definition:
\begin{equation}
	\sqrt{\eta} \varepsilon ^{\theta\phi}=  r^2 \sin\theta \varepsilon
	^{\theta\phi}=1\,.
\end{equation}
Let us introduce the denoting: $\lambda = \sqrt{\gamma}$ and reformulate the
right hand side of the thesis:
\begin{equation}
	\label{prawamoment}
	\begin{aligned}
		\int\limits_{S(r)}P ^{i}{}_{j} \mathcal{R}^j_z dS_i
		%= \int\limits_{S(r)} \lambda \left(
		%K ^{r}{}_{A} - \delta ^{r}{}_{A}K\right) r^2 \varepsilon ^{AB}
		%(\cos\theta) _{,B}
		&=
		-\int\limits_{S(r)} \lambda
		K ^{r}{}_{A}   r^2 \varepsilon ^{AB}
		(\cos\theta) _{,B}
		=\int\limits_{S(r)} (\lambda
		K ^{r}{}_{A}   r^2 \varepsilon ^{AB})_{,B}
		\cos\theta = \\ &=
		\int\limits_{S(r)} (\lambda
		K_{rA}   r^2 \varepsilon ^{AB})_{,B}
		\cos\theta
		=\int\limits_{S(r)} r^2(\lambda
		K_{rA}    \varepsilon ^{AB})_{||B}
		\cos\theta = \\ &=
		\int\limits_{S(r)} r^2\lambda
		\varepsilon ^{AB}K_{rA||B}
		\cos\theta \,.
	\end{aligned}
\end{equation}
First, we performed integration by parts relative to the index $B$, then we left the index to get $K _{rA}$, then we converted the partial derivative to covariant one (because the object being differentiated is the vector density on
sphere), in the next step we  pulled out the antisymmetric tensor and the volume element before the derivative. \\
For the left hand side of the thesis:
\begin{equation}
	\label{lewamoment}
	\begin{aligned}
		\int_{S(r)} B ^{i}{}_{j}\mathcal{K}^j_z dS_i &= \int_{S(r)} \lambda B ^{r}{}_{j}
	\bar{\mathcal{K}}_z^j = \int_{S(r)} \lambda \varepsilon ^{lsr}K _{jl|s}
	\bar{\mathcal{K}}_z^j = \int_{S(r)} \lambda \varepsilon ^{ABr} K _{rA|B}
	r^2\cos\theta = \\ &= \int_{S(r)} r^2\lambda \varepsilon ^{AB} K
	_{rA||B}\cos\theta\,.
	\end{aligned}
\end{equation}
By comparing the formulae (\ref{prawamoment}) and (\ref{lewamoment}) we get the thesis.
\qed

\vskip 2mm
\noindent Note that from the mentioned theorems it follows that in the case of ``electromagnetic'' charges, linear momentum generators are rotations, and  angular momentum generators are proper conformal transformations.

\noindent Let us examine the properties of transformation of charges related to the linear and angular momentums relative to the shift of the coordinate system by a fixed vector. Consider two
Cartesian coordinate systems $x^i$, $y^i$ shifted by a fixed vector
$a^i$. The relationship $y^i = x^i + a^i$ is satisfied.\\
\noindent In the case of classical mechanics, we received that the momentum remains invariant under translation:
\begin{equation}
	\vec{p}(y) = m \vec{v}_y = m \vec{v}_x = \vec{p}(x)\,,
\end{equation}
where $\vec{p}(x)$ means ``classical'' linear momentum in the coordinate system $x^i$, and
$\vec{v}_x$ is the velocity of the body of mass $m$ in the system $x$.\\
\noindent For the ``classical'' angular momentum $\vec{j}$:
\begin{equation}
	\label{angularmomentumclassical}
	\vec{j}(y) = \vec{r}_y \times \vec{p}(y) = (\vec{r}_x+\vec{a})\times
	\vec{p}(x) = \vec{r}_x \times \vec{p}(x) + \vec{a} \times \vec{p}(x)=
	\vec{j}(x) + \vec{a} \times \vec{p}(x)\,.
\end{equation}
The same transformational properties have linear and angular momentums for ADM formulation. To demonstrate this, let us recall the expressions for conformal Killing vectors in the shifted coordinate system:
\begin{equation}
	\label{translationTransf}
\mathcal{T}_k(x)=\mathcal{T}_k(y)\,,
\end{equation}
\begin{equation}
\mathcal{S}(y) = \mathcal{S}(x)+\mathcal{T}_ka^k\,,
\end{equation}
\begin{equation}
	\label{rotationTransf}
\mathcal{R}_k(y)=\mathcal{R}_k(x)+\varepsilon_k{}^{ij}a_i\mathcal{T}_j(x)\,,
\end{equation}
\begin{equation}
	\label{transformationConformal}
\mathcal{K}_k(y)=\mathcal{K}_k(x)+a_k\mathcal{S}(x)+a^l\mathcal{R}_{kl}(x)+\mathcal{T}_l(x)(a_ka^l-\frac{1}{2}a^ia_i\delta^l_k)\,,
\end{equation}
where
$\mathcal{R}_{kl}:=x_k\partial_l-x_l\partial_k\,.$
From the equation (\ref{translationTransf}) we have:
\begin{equation}
	P ^{i}{}_{j} (\mathcal{T}_k)^j(y) =
	P ^{i}{}_{j} (\mathcal{T}_k)^j(x)\,,
\end{equation}
which reproduces the classic transformational law for the momentum, while from the equation
(\ref{rotationTransf}):
\begin{equation}
	P ^{i}{}_{j} (\mathcal{R}_k)^j(y) =
	P ^{i}{}_{j} \left[  (\mathcal{R}_k)^j(x)+
	\varepsilon_k{}^{ml}a_m(\mathcal{T}_l)^j(x)\right]=
	P ^{i}{}_{j} (\mathcal{R}_k)^j(x) +
	\varepsilon_k{}^{ml}a_m P ^{i}{}_{j}(\mathcal{T}_l)^j(x)
\end{equation}
we recreate the transformation (\ref{angularmomentumclassical}).\\
Let us perform analogical reasoning for the charges defined by means of
magnetic part, first for the momentum:
\begin{equation}
	B ^{i}{}_{j} (\mathcal{R}_k)^j(y) =
	B ^{i}{}_{j} (\mathcal{R}_k)^j(x) +
	\varepsilon_k{}^{ml}a_m B ^{i}{}_{j}(\mathcal{T}_l)^j(x)\,.
\end{equation}
We see that the analogy with classical transformational law requires the assumption that the charge constructed from the magnetic part and the translational generator field is zero. Let us note that this is also one of the assumptions in the theorem
\ref{theoremangularmomentum}. \\
For the angular momentum, we use the equation (\ref{transformationConformal}):
\begin{equation}
	\begin{aligned}
		B ^{i}{}_{j} (\mathcal{K}_k)^j(y) &=
	B ^{i}{}_{j} (\mathcal{K}_k)^j(x) +
	B ^{i}{}_{j} a_k\mathcal{S}^j(x) +
	B ^{i}{}_{j} a^l (\mathcal{R} _{kl})^j(x) + \\
	&+B ^{i}{}_{j} (\mathcal{T}_l)^j(x) (a_ka^l-\frac{1}{2}a^ma_m\delta
	^{l}{}_{k})\,.
\end{aligned}
\end{equation}
Comparing with the transformation law (\ref{angularmomentumclassical}) requires, as previously, the assumption $Q(B,\mathcal{T}_k)=0$ and additionally
$Q(B, \mathcal{S})=0$.

%TODO ze ladunek Q(B,S) to przestrzenie taub-nut o nietrywialnej topologii, ladunek
%nazywany jest masa dualna

\section{Schwarzschild--de Sitter spacetime}
\label{chapterSchwarzschild}

We apply the concepts and theorems introduced in section
\ref{ladunkiKwazilokalne} to the analysis of conserved quantities (charges) in
 Schwarzschild--de Sitter spacetime. We will consider two families of
foliations with spatial hypersurfaces $\Sigma$. The first of them, denoted
$\Sigma_s$, corresponds to the surfaces of constant time (i.e. the coordinate appearing in the standard form of Schwarzschild metric)
$t_s=\textrm{const.}$, the second (denoted $\Sigma_p$) will be a hypersurface foliation with a flat inner geometry. In both cases
we calculate the electrical and magnetic parts from the initial data $(\gamma _{ij}, K
_{ij})$ and the conserved quantity corresponding to the mass. The obtained results are compared with the ADM mass.
We will show that the mass defined as the integral of the electrical part with the appropriate conformal Killing tensor is proportional to the parameter $M$ appearing in the metric and does not depend on the cosmological constant.

\subsection{The constant time hypersurfaces}

Consider the Schwarzschild--de Sitter metric with the standard variables
$(t_s,r,\theta,\phi)$, whose linear element is given by the formula:
\begin{equation}
	\label{schwarzschilddesitter}
	ds^2 = -f(r)dt_s^2+ \frac{dr^2}{f(r)}+r^2d\Omega^2\,,
\end{equation}
where $f(r)=1-\frac{2M}{r}-br^2$, and we assume that $f(r)>0\,.$ %\\
Parameter $b$ is a scaled cosmological constant $b=\frac{\Lambda}{3}$;
because we are considering the de Sitter spacetime, we assume $b\geq0$. In the special case of $b=0$, we get the Schwarzschild metric, and for $M=0$ the de Sitter metric.\\
If we assume that $b>0$, $M>0$, $r>0$, then the condition $f(r)>0$ takes the form:
\begin{equation}
	1-\frac{2M}{r}-br^2 > 0 \iff r - 2M - br^3 >0\,.
\end{equation}
The function $g(r):=rf(r)$ is a third degree polynomial, in general it can have three roots. Note that $g(0)=-2M < 0$ and
$\lim\limits_{r \to -\infty} g(r)=\infty$, hence $g(r)$ is assumed to have a
zero for a negative $r$. We assume that the values of the parameters $b$ and $M$ allow for the existence of two (different) real, positive roots $g(r)$, we denote them $r_-$, $r_+$, where $r_-<r_+$, then $r \in (r_-,r_+) \Rightarrow f(r)>0$
in accordance with our assumptions.\\
We will examine the foliation with spatial hypersurfaces
$\Sigma_s:t_s=\textrm{const}$. From the equation (\ref{schwarzschilddesitter}), we see that three-dimensional Riemannian metric induced on $\Sigma_s$  has the following form:
\begin{equation}
	\label{3dSchw}
	ds_3^2=\frac{dr^2}{f(r)}+r^2d\Omega\,.
\end{equation}
We can check by direct calculations that the Cotton tensor for the metric
(\ref{3dSchw}) is equal to zero, which (in three dimensions) is a necessary and sufficient condition for conformal flatness. Using lemma \ref{CKVflatlemma}
we see that $\Sigma_s$ has a full set of ten basic conformal Killing vectors. Let us try to find the conformal factor $\Omega$ and the appropriate coordinate change, which transforms the metric (\ref{3dSchw}) into the form $\Omega^2\eta _{ij}$, where $\eta _{ij}$ is a flat three-dimensional metric. We write the metric in the form:
\begin{equation}
	ds_3^2=\frac{dr^2}{f(r)}+r^2d\Omega = r^2 \left[ \left( \frac{dr}{r\sqrt{f(r)}}
	\right)^2+d\Omega \right]\,.
\end{equation}
Next, we define a new variable $x$ such that $dx= \frac{dr}{r\sqrt{f(r)}}$, so the
metric in new variables is:
\begin{equation}
	ds_3^2 = r^2 \left[ dx^2 + d\Omega \right]\,.
\end{equation}
Introducing $x=\log R$, where it appears that:  $dx = \frac{dR}{R}$, the metric in variables $(R,\theta,\phi)$
takes the conformally flat form:
\begin{equation}
	ds_3^2 = \frac{r^2(\log R)}{R^2}\left[ dR^2 + R^2d\Omega \right]\,.
\end{equation}
Comparing the formulae defining the new variables $R$ and $x$ we get the first order differential equation, which allows to express a new variable $R$ by the variable $r$:
\begin{equation}
	\frac{dR}{R}=\frac{dr}{r\sqrt{f(r)}} \quad\Rightarrow\quad \log R = \int
	\frac{dr}{r \sqrt{1-\frac{2M}{r}-br^2}}\,.
\end{equation}
The solution in the general case of $b>0$ is an elliptical function. In the special case of $b=0$ we get an equation whose solution is expressed by elementary functions, calculations lead to the reconstruction of the known form of Schwarzschild metric in isotropic coordinates $(t_s,\bar{r},\theta,\phi)$:
\begin{equation}
g_{\mu \nu} dx^\mu dx^\nu=-\left(\frac{1-M/2\bar{r}}{{1+M/2\bar{r}}} \right)^2
dt^2 +\left(1+\frac{M}{2\bar{r}} \right)^4 \left[d \bar{r}\:{}^2+\bar{r}\:{}^2
\left(d\theta^2+\sin^2\theta d\phi^2\right)\right]\,,
\label{newschwarz}
\end{equation}
where the new variable $\bar{r}$ is defined by:
\begin{equation}
	r=\bar{r}(1+\frac{M}{2\bar{r}})^2\,.
\end{equation}
In the general case of $b \geq 0$, we can not provide an explicit formula for $R$. We can, however, find the radial conformal Killing field, whose integral after contraction with the electrical part will be equal to the charge associated with the mass.\\
As before, consider the three-dimensional Riemannian space described in the coordinates $(r,\theta,\phi)$ and equipped with the metric of the form:
	\begin{equation}
		\label{CKV0}
		\gamma _{ij} = \textrm{diag}\left[\frac{1}{f(r)}, r^2,
		r^2\sin^2\theta\right]\,.
	\end{equation}
	We  look for a conformal Killing field $X^i$, which has a component only in the radial direction $r$. Let us write the equation for the conformal Killing field:
	\begin{equation}
	\label{CKV1}
		X _{i|j} + X _{j|i} = \lambda \gamma _{ij}\,.
	\end{equation}
	The equation must be satisfied in particular for $i=j=\theta$, hence:
	\begin{equation}
		\label{CKV2}
		2 X _{\theta,\theta}-2\Gamma ^{r}{}_{\theta\theta}X _{r}=\lambda\gamma
		_{\theta\theta}=\lambda r^2\,.
	\end{equation}
	By assumption, $X$ has a component only in the radial direction, so $X_\theta=0$
	and $X_{\theta,\theta}=0\,.$ %\\
	From the equation (\ref{CKV2}) we can recover the form of the function $\lambda$:
	\begin{equation}
		\lambda = -\frac{2}{r^2}\Gamma ^{r}{}_{\theta\theta}X _{r}\,.
	\end{equation}
	Then, write the equation (\ref{CKV1}) for $i=j=r$ and use the above form of the function $\lambda$:
	\begin{equation}
	\label{CKV3}
		2X _{r,r}-2\Gamma ^{r}{}_{rr}X _{r}=\lambda \gamma
		_{rr}=-\frac{2}{f(r)r^2}\Gamma ^{r}{}_{\theta\theta}X _{r}\,.
	\end{equation}
	The needed Christoffel symbols for the metric (\ref{CKV0}) are:
	\begin{equation}
		\Gamma ^{r}{}_{rr}=- \frac{f'(r)}{2f(r)}\,, \quad \; \quad \quad
		\Gamma ^{r}{}_{\theta\theta}=-rf(r)\,.
	\end{equation}
	The equation (\ref {CKV3}) becomes:
	\begin{equation}
		\frac{X _{r,r}}{X_r}=\frac{1}{r}-\frac{f'(r)}{2f(r)}\,,
	\end{equation}
	the solution is the function:
	\begin{equation}
		\log(|X _{r}|) = \log(|r|) - \frac{1}{2}\log(|f(r)|) +C = \log(
		\frac{r}{\sqrt{f}}) + C\,.
	\end{equation}
	Since we are interested in one particular solution, and by definition conformal Killing vectors are determined up to a multiplicative constant, we can skip the module and set $C=0$, hence:
	\begin{equation}
		\label{CKVmasySch}
		X_r = \frac{r}{\sqrt{f(r)}} \quad \Rightarrow \quad
		X^r = X_r \gamma^{rr} = r\sqrt{f(r)}\,.
	\end{equation}
	Note that the above reasoning for recovering the radial conformal Killing vector is independent on the form of the function $f(r)$, as long as we assume that the metric $\textrm{diag} \left(
	\frac{1}{f(r)},r^2,r^2\sin^2\theta \right)$ is conformally flat. The conditions for the function $f$ can be obtained in the general case by calculating the Cotton tensor and setting it to zero.

	\noindent Let us return to the considerations for Schwarzschild--de Sitter space with
	$f(r)=1-\frac{2M}{r}-br^2$. %\\
	We already know the form of the radial conformal Killing field, to calculate the mass we need an electrical part, which according to the theorem
	(\ref{theoremEBlambda}) can be expressed by the extrinsic curvature tensor and the three-dimensional Ricci tensor. Using the formula:
	\begin{equation}
		K _{ij}= \frac{1}{2N} \left( N _{i|j} + N _{j|i} - \partial_{t_s}\gamma _{ij} \right)
	\end{equation}
	and obtaining the lapse function $N$ and the shift vector from the form of the metric (\ref{schwarzschilddesitter}) we receive that the tensor of the extrinsic curvature $K _{ij}$ of hypersurface $\Sigma_s$  is identically equal to zero. According to the theorem (\ref{theoremEBlambda}) the magnetic part is zero, and the expression for the electrical part is reduced to:
	\begin{equation}
		\label{asdf}
			E _{i j} = -\overset{3}{R}{}_{i j} +\frac{2}{3}\Lambda
			\overset{3}{g}{}_{ij} =  -\overset{3}{R}
			_{ij} +2b \gamma_{ij}\,.
	\end{equation}
	The disappearance of the extrinsic curvature tensor causes the covariant divergence of the electrical and magnetic parts vanish (theorem
	\ref{divergenceTheorem}), and therefore charges constructed by contractions  $E$ or $B$ with conformal Killing vectors do not depend on the choice of the two-dimensional integration surface.\\
	Let us calculate the components of the electrical part.%\\
	The second derivative of the three-dimensional metric enables us to obtain the Riemann tensor, and after the contraction -- the Ricci tensor. We get the non-zero components of the three-dimensional Ricci tensor:
	\begin{equation}
		\overset{3}R _{rr} = \frac{2(M -br^3)}{r^2(2M-r+br^3)}\,,
	\end{equation}
	\begin{equation}
		\overset{3}R _{\theta \theta}= \frac{M+2br^3}{r}\,,
	\end{equation}
	\begin{equation}
		\overset{3}R _{\phi \phi} = \frac{M+2br^3}{r}\sin^2\theta\,.
	\end{equation}
	Next, we use the equation (\ref{asdf}). Non-zero components of the electrical part are the following:
	\begin{equation}
		\label{EdoMasy}
		E _{rr} = - \frac{2M}{r^2(2M-r+br^3)}\,,
	\end{equation}
	\begin{equation}
		E _{\theta \theta} = - \frac{M}{r}\,,
	\end{equation}
	\begin{equation}
		E _{\phi \phi} = - \frac{M}{r} \sin^2\theta\,.
	\end{equation}
	After raising one of the indices, we get quantities independent on the cosmological constant:
	\begin{equation}
		E ^{r}{}_{r} = \frac{2M}{r^3}\,,
	\end{equation}
	\begin{equation}
		E ^{\theta}{}_{\theta} = -\frac{M}{r^3}\,,
	\end{equation}
	\begin{equation}
		E ^{\phi}{}_{\phi} = -\frac{M}{r^3}\,.
	\end{equation}
	Using (\ref{EdoMasy}) and the form of a conformal Killing field
	(\ref{CKVmasySch}) we can calculate the charge responsible for mass:
\begin{equation}
	\begin{aligned}
		Q(E,\mathcal{S}) &= \int_{S(r)} E ^{r}{}_{r} X^r \sqrt{\gamma} d\theta d\phi = \int_{S(r)}
	\frac{2M}{r^3}r\sqrt{f(r)} \frac{1}{\sqrt{f(r)}}r^2\sin\theta d\theta
	d\phi=\\ &=8\pi M\,.
	\end{aligned}
\end{equation}

\noindent The obtained result is comparable with the ADM mass. We will use formulas from
\cite{3} to calculate ADM mass relative to any vector field $X$
and any reference space. Let $\gamma _{ij}$ denote the metric at
$\Sigma_s$, in the case of $b=0$ the reference spacetime is flat Minkowski space, and for $b>0$ the de Sitter spacetime. The reference metric is denoted by $\beta _{ij}$. Assuming the above, the ADM mass is expressed by the formula:
\begin{equation}
	m _{\textrm{ADM}} = \frac{1}{16\pi} \int_{S(r)} \left( \mathbb{U}^r +
	\mathbb{V}^r \right) \,,
\end{equation}
where:
\begin{equation}
	\mathbb{U}^i(V) = 2 \sqrt{\textrm{det}\gamma} \left[ V
	\gamma^{i[k}\gamma^{j]l}\bar{D}{}_{j}\gamma_{kl}
	+ D ^{[i}V \gamma ^{j]k}e _{jk}\right]\,,
\end{equation}
\begin{equation}
	\mathbb{V}^l(Y) = 2 \sqrt{\textrm{det}\gamma} \left[ (P ^{l}{}_{k}-\bar{P} {}^{l}{}_{k})Y^k -
	\frac{1}{2}Y^l \bar{P}{} ^{mn}e _{mn}+\frac{1}{2}Y^k \bar{P}{}
	^{l}{}_{k}\beta^{mn}e _{mn}\right]\,,
\end{equation}
\begin{equation}
	e _{ij}:=\gamma _{ij}-\beta _{ij}\,,
\end{equation}
\begin{equation}
	P ^{lk} := \gamma ^{lk}\textrm{tr}_\gamma K - K
	^{lk}\,,\quad\quad\textrm{tr}_\gamma K:=\gamma ^{lk}K
	_{lk}\,,
\end{equation}
similarly for objects defined in the background space:
\begin{equation}
	\bar{P} {}
	^{lk}:=\beta ^{lk}\textrm{tr}_\beta \bar{K} - \bar{K}{}
	^{lk}\,,\quad\quad\textrm{tr}_\beta\bar{K}:=
	\beta^{lk}\bar{K}{} _{lk}\,.
\end{equation}
Form of the field $X$:
\begin{equation}
	X = V n^\mu \partial_\mu + Y^k\partial_k = \frac{V}{N}\partial_0 + \left(Y^k -
	\frac{V}{N}N^k\right) \partial_k	\,.
\end{equation}
Because we count mass (energy), then $X=\partial_0$, so $V=N$ and $Y^k =
N^k$. In our case, i.e. the Schwarzschild--de Sitter spacetime and the surface $\Sigma_s$ of constant Schwarzschild time:
\begin{equation}
	\beta _{ij} = \textrm{diag} \left(\frac{1}{1-br^2},r^2,r^2\sin^2\theta
	\right)\,,
\end{equation}
\begin{equation}
	\gamma _{ij} = \textrm{diag} \left( \frac{1}{1-\frac{2M}{r}-br^2}, r^2,
	r^2\sin^2\theta \right)\,,
\end{equation}
hence:
\begin{equation}
	e _{ij} = \textrm{diag} \left( \frac{1}{1-\frac{2M}{r}-br^2}-
	\frac{1}{1-br^2},0,0 \right)\,.
\end{equation}
The extrinsic curvature is zero, therefore $P ^{k}{}_{l}=\bar{P} {} ^{k}{}_{l}=0
\Rightarrow \mathbb{V}^l(Y)=0\,.$\\
The expression for mass is reduced to:
\begin{equation}
	m_\textrm{ADM}=\frac{1}{8\pi}\int_{S(r)}\sqrt{\textrm{det}\gamma} \left[ N
	\gamma ^{r[k}\gamma ^{j]l}\bar{D}{}_{j}\gamma _{kl}
		+ D ^{[r}N\gamma ^{j]k}e _{jk}\right]\,.
\end{equation}
After calculations:
\begin{equation}
	m_\textrm{ADM} = M \,.
\end{equation}
Thus, the ADM mass and ''electromagnetic'' mass are equal (up to the normalization factor $8\pi$), both are independent on the radius $r$ and
the parameter proportional to the cosmological constant $b$.

%%%%%%%%%%%%%%%%%%%%%%%%%%%%%%%%%%%%%%%%%%%%%%%%%%%%%%%%%%%%%%%%%%%%%%%%%%%%%%%%
%                          HIPERPOWIERZCHNIE PLASKIE                           %
%%%%%%%%%%%%%%%%%%%%%%%%%%%%%%%%%%%%%%%%%%%%%%%%%%%%%%%%%%%%%%%%%%%%%%%%%%%%%%%%

\subsection{Flat hypersurfaces}
As in the previous section, let us consider the metric form:
\begin{equation}
	ds^2 = -f(r)dt_s^2+ \frac{dr^2}{f(r)}+r^2d\Omega^2\,,
\end{equation}
for $f=1-\frac{2M}{r}-br^2$, $f>0$, $b\geq0$, $M \geq 0$.\\
We will examine the foliation with the hypersurfaces $\Sigma_t$, on
which induced three-dimensional metric is flat. Let us define a new variable $t_p$:
\begin{equation}
	t_p := t_s - g(r) \quad \Rightarrow \quad dt_p = dt_s - g'(r)dr\,.
\end{equation}
The metric in new variables $(t_p,r,\theta,\phi)$ has the form:
\begin{equation}
\label{painlevemetric}
	\begin{aligned}
	ds^2 &= -f(r)(dt_p + g'(r)dr)^2 + \frac{dr^2}{f(r)}+ r^2\Omega^2= \\
	&= -f(r)dt_p^2 -2f(r)g'(r)drdt_p +
	\left(\frac{1}{f(r)}-f(r)[g'(r)]^2\right)
	dr^2 + r^2 \Omega^2\,.
	\end{aligned}
\end{equation}
We want to find such function $g(r)$, for which the spatial part of the above metric in the new variables $(t_p,r,\theta,\phi)$ is flat. From the form
(\ref{painlevemetric}) we see that the following equation must be satisfied:
	\begin{equation}
	\label{diffequationpainleve}
		\frac{1}{f(r)}-fg'(r){}^2=1 \quad \Rightarrow \quad g'(r) =
		\frac{\sqrt{1-f(r)}}{f(r)}\,.
	\end{equation}
	In the general case of $b\neq0$ the solution of the equation
	(\ref{diffequationpainleve}) for $ f(r) = 1- \frac{2M}{r}-br^2$ is not expressed by elementary functions. In the special case of $b=0$, that is for Schwarzschild spacetime, we get the equation:
\begin{equation}
	g'(r) = \frac{\sqrt{2Mr}}{r-2M}\,,
\end{equation}
whose solution is:
\begin{equation}
	g(r) = 2\sqrt{2Mr}+2M\log \left(
	\frac{\sqrt{r}-\sqrt{2M}}{\sqrt{r}+\sqrt{2M}} \right)\,,
\end{equation}
then the spacetime metric takes the Painleve--Gullstrand form:
\begin{equation}
\label{metricpainleve}
	ds^2 = -dt_p^2+\left( dr - \sqrt{\frac{2m}{r}}dt_p \right)^2 +
	r^2d\Omega^2\,.
\end{equation}
%The metric (\ref{metricpainleve}) is called the Painleve--Gullstrand metric. \\
%TODO dopisac czemu odpowiada i jaka jest interpretacja.

\noindent Note that in the general case of $b \neq 0$ we do not need an explicit formula for the function $g(r)$. To write the metric (\ref{painlevemetric}) you only need to know the derivative of $g(r)$:
\begin{equation}
		ds^2	= -f(r)dt_p^2 -2\sqrt{1-f(r)}drdt_p +
							dr^2 + r^2 \Omega^2\,.
\end{equation}
We can also write a metric using the assumed form of function $f(r)$:
%Jeżeli wykorzystamy ogólną założoną postać $f(r)$:
\begin{equation}
		ds^2	= -\left( 1-\frac{2 M}{r}-br^2 \right)dt_p^2 -2\sqrt{\frac{2
		M}{r}+br^2}\:drdt_p + dr^2 + r^2 \Omega^2\,.
\end{equation}
From the form of the metric we can read the lapse $N$ and the shift vector $N^i$:
\begin{equation}
	N=1\,,
\end{equation}
\begin{equation}
	N_r = -\sqrt{\frac{2M}{r}+br^2}\,, \quad\quad N_\theta = 0\,, \quad\quad
	N_\phi =0\, ,
\end{equation}
and then calculate the extrinsic curvature from the formula:
\begin{equation}
	K _{ij}= \frac{1}{2N} \left( N _{i|j} + N _{j|i} - \partial_{t_s}g _{ij}
	\right)\,.
\end{equation}
Non-zero components of the extrinsic curvature are:
\begin{equation}
	K _{rr} = \frac{M-br^3}{\sqrt{2Mr^3+br^6}}\,,
\end{equation}
\begin{equation}
	K _{\theta \theta} = -\sqrt{2Mr+br^4}\,,
\end{equation}
\begin{equation}
	K _{\phi \phi} = -\sqrt{2Mr+br^4}\sin^2\theta\,.
\end{equation}
The trace of extrinsic curvature is equal to:
\begin{equation}
	K = K _{ij}\gamma ^{ij} = - \frac{3(M+br^3)}{\sqrt{2Mr^3+br^6}}\,.
\end{equation}
Let us calculate the electrical part of Weyl tensor, the expression (\ref{lematElambda}) reduces to:
\begin{equation}
		E _{i j} = - K K _{i j} +
			K _{i k}K ^{k}{}_{j}+\frac{2}{3}\Lambda \gamma_{ij}
			=- K K _{i j} + K _{i k}K ^{k}{}_{j}+2b \gamma_{ij}\,.
\end{equation}
After the calculation, we obtain the following non-zero components of the electrical part of Weyl tensor:
%Niezerowe składowe części elektrycznej:
\begin{equation}
	E _{rr} = \frac{2M}{r^3}\,,
\end{equation}
\begin{equation}
	E _{\theta \theta}= - \frac{M}{r}\,,
\end{equation}
\begin{equation}
	E _{\phi \phi} = - \frac{M}{r}\sin^2\theta\,.
\end{equation}
The magnetic part is zero.\\
Because the metric induced on $\Sigma_p$ is flat, we have a full set of ten conformal Killing vectors. Note that the disappearance of the magnetic part causes that the three-dimensional covariant divergence of the electrical part vanishes (theorem \ref{divergenceTheorem}), and therefore the ``electric'' charges do not depend on the two-dimensional integration surface. In this case, due to symmetry, integration over the sphere is very simple.
Let us calculate the contraction of the electrical part with the basic conformal Killing vector $\mathcal{S}=r\partial_r$; by integrating this expression, we get the charge responsible for the mass:
\begin{equation}
	\label{massESpainleve}
	Q(E,\mathcal{S})=\int_{S(r)} E ^{i}{}_{j}\mathcal{S}^j dS_i = \int_{S(r)} E ^{r}{}_{r}
	r^3\sin\theta d\theta d\phi=8\pi M\,.
\end{equation}
%\vskip 5mm
\noindent We compare the obtained result with the ADM mass. As in the previous subsection, we define:
\begin{equation}
	e _{ij}:=\gamma _{ij}-\beta _{ij}\,,
\end{equation}
\begin{equation}
	\mathbb{U}^i(V) = 2 \sqrt{\textrm{det}\gamma} \left[ V \gamma ^{i[k}\gamma
	^{j]l}\bar{D}{}_{j}\gamma _{kl}
	+ D ^{[i}V\gamma ^{j]k}e _{jk}\right]\,,
\end{equation}
\begin{equation}
	\mathbb{V}^l(Y) = 2 \sqrt{\textrm{det}\gamma} \left[ (P ^{l}{}_{k}-\bar{P} {}^{l}{}_{k})Y^k -
	\frac{1}{2}Y^l \bar{P}{} ^{mn}e _{mn}+\frac{1}{2}Y^k \bar{P}{}
	^{l}{}_{k}\beta^{mn}e _{mn}\right]\,,
\end{equation}
%postać pola $X$:
\begin{equation}
	X = V n^\mu \partial_\mu + Y^k\partial_k = \frac{V}{N}\partial_0 + \left(Y^k -
	\frac{V}{N}N^k\right) \partial_k\,.
\end{equation}
We obtain the mass, therefore $X=\partial_0 \Rightarrow V=N$ and $Y^k=N^k$.%\\
$\mathbb{U}^l$ is zero, because in the case of foliations by flat hypersurfaces $\gamma _{ij} = \beta _{ij}$, where $e _{ij}=0$, and in the first part the covariant derivative $\bar{D}$ can be converted to $D$.\\
Let us calculate the canonical ADM momentum:
\begin{equation}
	P ^{ij} = \gamma ^{ij}K - K ^{ij} \Rightarrow P ^{i}{}_{j} = \delta ^{i}{}_{j}K -
	K ^{i}{}_{j}\,.
\end{equation}
After the calculation, we obtain that non-zero components of the ADM momentum are:
\begin{equation}
	\label{ADMmomfirst}
	P ^{r}{}_{r} = - \frac{2\sqrt{2M+br^3}}{r^{3/2}}\,,
\end{equation}
\begin{equation}
	P ^{\theta}{}_{\theta} = - \frac{M+2br^3}{r^{3/2}\sqrt{2M+br^3}}\,,
\end{equation}
\begin{equation}
	\label{ADMmomlast}
	P ^{\phi}{}_{\phi} = - \frac{M+2br^3}{r^{3/2}\sqrt{2M+br^3}}\,.
\end{equation}
Next, we calculate the ADM momentum for reference time and space. We assume that for $b=0$
the reference is Minkowski space, and for $b>0$ is de Sitter space.
Note that the background ADM momentum $\bar{P} ^{i}{}_{j}$ can be obtained from the
equations (\ref{ADMmomfirst})--(\ref{ADMmomlast}) by setting  $M=0$, hence we get:
\begin{equation}
	\bar{P} ^{r}{}_{r} = \bar{P} ^{\theta}{}_{\theta}= \bar{P}
	^{\phi}{}_{\phi}=-2\sqrt{b}\,.
\end{equation}
We raise the index in the translation vector with a (flat) three-dimensional metric, thus:
%indeks w shifcie podnosze metryka trojwymiarowa, czyli $N_r=N^r$
\begin{equation}
	N_r=N^r= -\sqrt{\frac{2M}{r}+br^2}\,.
\end{equation}
The only non-omitting element in the expression for ADM mass is:
\begin{equation}
	m_{\textrm{ADM}} = \frac{1}{16\pi} \int_{S(r)} 2\sqrt{\gamma}(P
	^{r}{}_{r}-\bar{P}{}^{r}{}_{r})N^r d\theta d\phi\,.
\end{equation}
After calculating the integral, we get:
\begin{equation}
	m_{\textrm{ADM}} =  2M+br^3-\sqrt{br^3(2M+br^3)} \,.
\end{equation}
Note that for Schwarzschild spacetime ($b=0$) we get $m
_{\textrm{ADM}} = 2M$, which is twice as much as the mass parameter in the metric. This is caused by too slow disappearance of the extrinsic curvature tensor. In the case of foliation with surfaces of flat internal geometry (Painleve--Gullstrand foliation), the tensor of the extrinsic curvature behaves like $r^{-3/2}$, and usually assumed assumption
when counting ADM mass is behavior like $r^{-3/2-\varepsilon}$, where $\varepsilon$
is strictly positive. The result (\ref{massESpainleve}) shows that the ``electromagnetic'' mass (at least in this case) has better properties,
because after dividing by the normalizing factor $8\pi$ accurately reproduces the parameter $M$ occurring in the metric, and is independent of the radius $r$ and the
cosmological constant $b$.

\section{Asymptotic charges}
\label{chapterAsymptotyczne}
%\section{Wprowadzenie}
In the section \ref{ladunkiKwazilokalne} we have given the definition of quasi-local charges and the conditions for obtaining strict rights of behavior. We assumed that the three-dimensional metric on $\Sigma$ is conformally flat and the covariant divergences of the electrical and magnetic parts disappear. We generalize the concepts introduced and provide the conditions for defining asymptotic ''electromagnetic'' charges. \\
Let us remind that the condition guaranteeing the independence of the value of the charge from the selection of the integration surface was the disappearance of the three-dimensional covariant divergence of the density contraction of the electrical (or magnetic) part with the conformal Killing vector.
\begin{equation}
	\label{asymptoticdivergence}
	\begin{aligned}
	(\sqrt{\gamma}E ^{ij}X _{j}) _{,i}&= (\sqrt{\gamma} E ^{ij} X _{j}) _{|i} =
		\sqrt{\gamma} E
	^{ij}X _{j|i} + \sqrt{\gamma}X^j(K\wedge B) _{j} =\\
	&=\sqrt{\gamma} E
	^{ij}X _{(j|i)} + \sqrt{\gamma}X^j(K\wedge B) _{j} =\\
	&=\sqrt{\gamma} E
	^{ij}(X _{(i|j)} -\frac{1}{3}X ^{k}{}_{|k}\gamma _{ij} )+
	\sqrt{\gamma}X^j(K\wedge B) _{j}\,.
	\end{aligned}
\end{equation}
Because  $E _{ij}$ is symmetric and traceless, in the second line we added symmetrization to the derivative of the field  $X$, and in the third line the expression proportional to the metric. Let us define a symmetric tensor $V _{ij}$ depending on the vector field $X$ and the metric $\gamma$:
\begin{equation}
	V _{ij}(X,\gamma) := X _{(i|j)} - \frac{1}{3} X ^{k}{}_{|k} \gamma _{ij}\,.
\end{equation}
Note that the definition of tensor $V _{ij}$  is the equation for conformal Killing vectors (in which all members have been moved to one side). If $X$ is a conformal Killing vector for $\gamma$ then $V _{ij}(X,\gamma)=0$.
Let us use $V$ to write the divergence (\ref{asymptoticdivergence}):
\begin{equation}
	(\sqrt{\gamma}E ^{ij} X _{j}) _{,i} = \sqrt{\gamma} E ^{ij} V _{ij}(X,\gamma)
	+\sqrt{\gamma} X^j (K \wedge E) _{j}\,.
\end{equation}
Consider the spatial hypersurface $\Sigma$ equipped with a conformally flat metric $\gamma$. According to the lemma \ref{CKVflatlemma} we have ten basic conformal Killing vectors $X$ for which
$V _{ij}(X,\gamma)=0$. Define asymptotic charges:
\begin{equation}
	Q_{as}(E,X) := \lim_{r \to \infty} \int_{S(r)} E ^{i}{}_{j}X^j dS_i\,,
\end{equation}
\begin{equation}
	Q_{as}(B,X) := \lim_{r \to \infty} \int_{S(r)} B ^{i}{}_{j}X^j dS_i\,.
\end{equation}
The boundaries in the definitions will be finite if the appropriate divergences are integrable at infinity, i.e.
%do zera
\begin{equation}
	\label{asymptotycznywarunekE}
	\sqrt{\gamma} X^j (K \wedge E) _{j}=O(r^{-1-\varepsilon})\,,
\end{equation}
\begin{equation}
	\label{asymptotycznywarunekB}
	\sqrt{\gamma} X^j (K \wedge B) _{j}=O(r^{-1-\varepsilon})\,,
\end{equation}
where $\varepsilon>0$. \\
Note that the basic conformal Killing vectors have different asymptotic relative to $r$,
using (\ref{CKVT})--(\ref{CKVS}) we have:
\begin{itemize}
	\item $\mathcal{T}_k = O(1)\,,$
	\item $\mathcal{R}_k = O(r)\,,$
	\item $\mathcal{S} = O(r)\,,$
	\item $\mathcal{K}_k = O(r^2)\,.$
\end{itemize}
Generators of proper conformal transformations behave like $r^2$,
therefore charges defined with $\mathcal{K}_{\mathbf k}$ (e.g. angular momentum)
impose the strongest conditions on the asymptotic of other objects found in the equations
(\ref{asymptotycznywarunekE})--(\ref{asymptotycznywarunekB}). By assumption, we consider asymptotically flat spaces, therefore the conformal factor of the metric tends to unity in spatial infinity, i.e.
$\sqrt{\gamma}=O(r^2)$. Analyzing the equations
(\ref{asymptotycznywarunekE})--(\ref{asymptotycznywarunekB}) we can give the minimum asymptotic needed to define angular momentum and center of mass:
\begin{equation}
	(K \wedge E) _{j}=O(r^{-5-\varepsilon})\,,
\end{equation}
\begin{equation}
	(K \wedge B) _{j}=O(r^{-5-\varepsilon})\,.
\end{equation}
Analogously for mass and momentum:
\begin{equation}
	(K \wedge B) _{j}=O(r^{-4-\varepsilon})\,,
\end{equation}
\begin{equation}
	(K \wedge E) _{j}=O(r^{-4-\varepsilon})\,.
\end{equation}
The above conditions are sufficient if we assume that the metric $\gamma _{ij}$
is conformally flat, which guarantees the existence of (strict) conformal Killing fields. \\
We can also consider surfaces $\Sigma$, whose metric is only asymptotically conformally flat. We will now prove the theorem in which we will assume that the metric is of the form:
\begin{equation}
	\gamma _{ij} = \left( 1+ \frac{M}{2r} \right)^4 \left( \eta _{ij} + h _{ij}
	\right)\,,
\end{equation}
where $\eta _{ij}$ is a flat metric, and $h _{ij}$ is sufficiently small at infinity.
The metric of such a form corresponds to slightly disturbed Schwarzschild spacetime
(see Schwarzschild metric expressed in isotropic variables
(\ref{newschwarz})).

\begin{theorem}
	\label{asymptotic}
	Let the metric on the hypersurface $\Sigma$ be in the form:
	%Zakładamy, że riemannowska metryka na hiperpowierzchni $\Sigma$ jest postaci:
	\begin{equation}
		\gamma _{ij} = \left( 1+ \frac{M}{2r} \right)^4 \left( \eta _{ij} + h
		_{ij} \right)\,,
	\end{equation}
	where $\eta _{ij}$ is a three-dimensional Euclidean metric.%\\
	We assume the following behavior of presented objects at $r \rightarrow \infty\;:$
	\begin{equation}
		h _{ij} =O\left( r^{-1-\varepsilon}\right)\,,
	\end{equation}
	\begin{equation}
		(\Gamma _{\eta+h}) ^{a}{}_{ij} = O\left( r^{-2-\varepsilon} \right)\,,
	\end{equation}
	\begin{equation}
		K _{ij} =O\left( r^{-3}\right)\,,
	\end{equation}
	\begin{equation}
		E _{ij} =O\left( r^{-3}\right)\,,
	\end{equation}
	\begin{equation}
		B _{ij} =O\left( r^{-3}\right)\,,
	\end{equation}
	where $\varepsilon >0\,.$ \\
	Then there is a finite limit at $r \rightarrow
	\infty$ for asymptotic charges responsible for mass, center of mass, linear momentum and angular momentum.
\end{theorem}
\noindent\textit{Proof:}\;
The aim of the proof is to show that at $r \rightarrow \infty$
divergences given by equations (\ref{asdivergenceE}), (\ref{asdivergenceB}) are integrable at infinity.
\begin{equation}
	\label{asdivergenceE}
	(\sqrt{\gamma}E ^{ij}X _{j}) _{,i}= (\sqrt{\gamma} E ^{ij} X _{j}) _{|i} = \sqrt{\gamma} E
	^{ij}V _{ij}(g,X) + \sqrt{\gamma}X^j(K\wedge B) _{j}\,,
\end{equation}
\begin{equation}
	\label{asdivergenceB}
	(\sqrt{\gamma}B ^{ij}X _{j}) _{,i}= (\sqrt{\gamma} B ^{ij} X _{j}) _{|i} = \sqrt{\gamma} B
	^{ij}V _{ij}(g,X) - \sqrt{\gamma}X^j(K\wedge E) _{j}\,,
\end{equation}
where $X$ is one of the vector fields belonging to the set $\{ \mathcal{S},
\mathcal{T}_k,
\mathcal{R}_k, \mathcal{K}_k\}$. We have:
\begin{equation}
	\mathcal{S}=O( r) \quad, \quad \mathcal{T}_k = O(1) \quad,\quad \mathcal{R}_k =O( r) \quad , \quad
	\mathcal{K}_k =O( r^2)\,.
\end{equation}
Because  $\mathcal{S}$, $\mathcal{R}_{\mathbf k}$ and $\mathcal{T}_{\mathbf k}$ have a weaker asymptotic than
$\mathcal{K}_{\mathbf k}$, it is sufficient to prove this statement for $X =O( r^2)$.\\
The minimum condition for ensuring integrability at infinity is that the asymptotic decay is better than $r^{-1-\varepsilon}$ in the limit $r \rightarrow \infty$, where
$\varepsilon>0$. \\
Let us assume that the field $X$ behaves at infinity like $r^2$ and check the asymptotic of the members containing the extrinsic curvature:
\begin{equation}
	\sqrt{\gamma} X^j(K \wedge B)_{j} =O(r^2)\cdot O(r^2) \cdot O(r^{-3}) \cdot
	O(r^{-3})=O(r^{-2})\,,
\end{equation}
\begin{equation}
	\sqrt{\gamma} X^j(K \wedge E)_{j} =O(r^2)\cdot O(r^2) \cdot O(r^{-3}) \cdot
	O(r^{-3})=O(r^{-2})\,.
\end{equation}
Thus, both expressions are integrable at infinity.
Let us examine expressions containing tensor $V _{ij}$. By assumption, the metric $\gamma$
is of the form: conformal factor times a small correction to a flat metric. The tensor $V$ behaves in conformal transformation as follows:
\begin{equation}
	V _{ij}\left(\Omega^4 g, X\right) = \Omega^4 V _{ij}(g, X)\,,
\end{equation}
where $g$ is any metric. In the case under consideration for $\Omega =
1+\frac{M}{2r}$:
\begin{equation}
	V _{ij}\left(\Omega^4(\eta+h),X\right) = \Omega^4 V _{ij}(\eta+h,X)\,.
\end{equation}
Let us use the definition of tensor $V$:
\begin{equation}
	\label{Vequation}
	V(\eta+h,X) = X _{(i,j)}-\Gamma ^{m}{}_{ij}X_m - \frac{1}{3} \left( X
	^{k}{}_{,k}+\Gamma ^{k}{}_{mk}X^m \right)(\eta _{ij}+h _{ij})\,.
\end{equation}
We assume that the field $X$ is a conformal Killing vector for a flat, three-dimensional metric $\eta$, thus the following equation is satisfied:
\begin{equation}
	X _{(i,j)}-\frac{1}{3}X ^{k}{}_{,k}\eta _{ij}=0\,.
\end{equation}
Hence (\ref{Vequation}) is equivalent:
\begin{equation}
	V(\eta+h,X) = -\Gamma ^{m}{}_{ij}X_m-\frac{1}{3}X ^{k}{}_{,k}h
	_{ij}-\frac{1}{3}\Gamma ^{k}{}_{mk}X^m(\eta _{ij}+h _{ij})\,.
\end{equation}
The Christoffel symbols appearing in the above formula come from the metric $\eta
+h$, we assume that their asymptotic behavior is at $r\rightarrow \infty$.
Assume that $X^i=O(r^2)$, hence:
\begin{equation}
	V \left( \Omega^4(\eta+h),X \right) = \Omega^4 V (\eta+h,X) =
	O(r^{-\varepsilon})\,.
\end{equation}
We have the following asymptotic of objects containing the tensor $V _{ij}$ in the divergences (\ref{asdivergenceE})--(\ref{asdivergenceB}):
\begin{equation}
	\sqrt{\gamma} E
	^{ij}V _{ij}(g,X) = O(r^2)\cdot O(r^{-3})\cdot O(r^{-\varepsilon}) = O(r
	^{-1-\varepsilon})\,,
\end{equation}
\begin{equation}
	\sqrt{\gamma} B
	^{ij}V _{ij}(g,X) = O(r^2)\cdot O(r^{-3})\cdot O(r^{-\varepsilon}) = O(r
	^{-1-\varepsilon})\,.
\end{equation}
Finally, we get that divergences (\ref{asdivergenceE}),
(\ref{asdivergenceB}) are integrable at infinity. Defined by equations:
\begin{equation}
	Q_{as}(E,X) := \lim_{r \to \infty} \int_{S(r)} E ^{i}{}_{j}X^j dS_i\,,
\end{equation}
\begin{equation}
	Q_{as}(B,X) := \lim_{r \to \infty} \int_{S(r)} B ^{i}{}_{j}X^j dS_i\,,
\end{equation}
asymptotic charges have a limit at $r \rightarrow \infty$, where $X^i$ are conformal Killing fields for the flat metric $\eta$ (and at the same time asymptotic Killing fields for the considered metric $\gamma$).
\qed

\section*{Summary}
\addcontentsline{toc}{chapter}{Summary}
We have shown that the method of defining four-dimensional gravity charges using conformal Yano--Killing tensors can be extended and used to define well-defined charges in a (3 + 1) decomposition.
The construction presented here enables one to calculate twenty local charges in terms of initial data $(\gamma _{ij}, K _{ij})$ on a conformally flat spatial hypersurface $\Sigma$ immersed in a spacetime satisfying the Einstein vacuum equations with a cosmological constant. Such specified charges are preserved (analogical to the Gaussian law), if the corresponding products of the extrinsic curvature with the electrical and magnetic parts disappear, in particular the condition of vanishing tensor of the extrinsic curvature is sufficient. Also in the case of ``purely magnetic'' and ``purely electrical'' spaces, all charges are preserved.\\
In addition, we have proved theorems explaining relation between linear momentum and angular momentum (defined as the contractions of the magnetic part of Weyl tensor with appropriate conformal Killing vectors) and the traditional ADM linear momentum and the ADM angular momentum. \\
The analysis performed for the Schwarzschild--de-Sitter spacetime showed that the mass defined as the contraction of the electrical part of the Weyl tensor with the scaling generator may have better properties than the ADM mass. In the proposed example of foliation with surfaces with flat internal geometry, the mass calculated with conformal Killing fields was not dependent on radius or cosmological constant (as opposed to ADM mass).\\
The thesis formulated in the last section shows that in certain specific cases we can use the proposed method to define asymptotic charges (even if the metric on spatial surfaces is not conformally flat).

\vspace{0.5 cm}

	{\noindent \sc Acknowledgements} This work was supported in part by Narodowe Centrum Nauki (Poland) under Grant No. 2016/21/B/ST1/00940.

%1) ze z CYK tensorow w rozkladzie 3+1 wychodza konforemne wektory killinga, ze
%mozna zdefiniowac w ten sposob ladunki (20), ktore daja sie dobrze zdefiniowac w
%pewnej klasie czasoprzestrzeni, mozna je wyliczyc z danych poczatkowych
%czasoprzestrzeni prozniowych ze stala kosmologiczna,
%2) pokazalismy ze pewne ladunki e-mag sa zwiazane z pedem i momentem pedu adm,
%3) podalismy przyklad czasoprzestrzeni w ktorych masa elektromagnetyczna ma
%lepsze wlasnosci niz masa adm
%4) klase przestrzeni w ktorych dobrze zdefiniowane sa ladunki asymptotyczne
%()

\appendix

\section{(3 + 1) Decomposition}
%%%%%%%%%%%%%%%%%%%%%%%%%%%%%%%%%%%%%%%%%%%%%%%%%%%%%%%%%%%%%%%%%%%%%%%%%%%%%%%%
%                                 Gauss-Codazzi                                 %
%%%%%%%%%%%%%%%%%%%%%%%%%%%%%%%%%%%%%%%%%%%%%%%%%%%%%%%%%%%%%%%%%%%%%%%%%%%%%%%%
%%%%%%%%%%%%%%%%%%%%%%%%%%%%%%%%%%%%%%%%%%%%%%%%%%%%%%%%%%%%%%%%%%%%%%%%%%%%%%%%
%                                    Gauss                                     %
%%%%%%%%%%%%%%%%%%%%%%%%%%%%%%%%%%%%%%%%%%%%%%%%%%%%%%%%%%%%%%%%%%%%%%%%%%%%%%%%
\subsection{Basic dependencies}
The four-dimensional metric is expressed by the lapse function $N$ and the shift vector ($N^m$) as follows:
\begin{equation}
\begin{bmatrix} g_{00}&g_{0k}\\g_{i0}&g_{ik} \end{bmatrix}=\begin{bmatrix} (N_s
N^s-N^2)& N_k \\ N_i & \gamma_{ik} \end{bmatrix} \, .
\end{equation}
Reverse metric:
\begin{equation}
\begin{bmatrix} g^{00}&g^{0m}\\g^{k0}&g^{km} \end{bmatrix}=\begin{bmatrix}
-(1/N^2)& (N^m/N^2)\\ (N^k/N^2) & (\gamma^{km}-N^kN^m/N^2) \end{bmatrix}\,.
\end{equation}
Unit time normal vector:
\begin{equation}
n_\mu=(-N,0,0,0), \qquad n^\mu=[(1/N),-(N^m/N)]\,.
\end{equation}
Relationship between the three-dimensional metric $\gamma$ and a four-dimensional metric $g$:
\begin{equation}
g^{\mu \nu}+n^\mu n^\nu = \gamma^{\mu \nu}, \qquad g_{\mu \nu}+n_\mu n_\nu =
\gamma_{\mu \nu}\,.
\end{equation}
Volume element:
\begin{equation}
	 \sqrt{-g}\;\textrm{d}x^0\textrm{d}x^1\textrm{d}x^2\textrm{d}x^3=N\;\sqrt{\gamma}\;\textrm{d}t\;\textrm{d}x^1\textrm{d}x^2\textrm{d}x^3\,.
\end{equation}
We will now derive the Gauss-Codazzi equations that form the basis of the (3 + 1) decomposition (\cite{formalism}). They allow to distribute the Riemann tensor of four-dimensional spacetime into objects associated with the spatial hypersurface $\Sigma$, tensor of the extrinsic curvature $K _{ij}$ and the three-dimensional Riemann tensor derived from the metric
$\gamma _{ij}$ induced on $\Sigma$.

\subsection{Gauss relations}
\label{sectionGauss}
Consider the formula (\ref{defRiem}), which defines the three-dimensional Riemann tensor, responsible for the non-commutation of covariant derivatives $D$ on the hypersurface $\Sigma$ provided with the metric $\gamma$:
\begin{equation}
\label{defRiem}
	(D _{i}D _{j}-D_jD_i)v^k = R ^{k}{}_{lij}v^l\,.
\end{equation}
The four-dimensional equivalent of the above formula can be written as:
\begin{equation}
\label{gauss0}
	(D _{\alpha}D _{\beta}-D_\beta D_\alpha)v^\gamma = R ^{\gamma}{}_{\mu \alpha
	\beta}v^\mu\,,
\end{equation}
where $v$ is a vector field tangent to $\Sigma$. Let us use the relationship
$\mathbf{DT}=\gamma^*\mathbf{\nabla T}$ connecting a three-dimensional covariant derivative $D$ with a four-dimensional covariant derivative $\nabla$:
\begin{equation}
	D _{\alpha}D _{\beta}v ^\gamma = D _{\alpha}(D _{\beta}v^\gamma) = \gamma
	^{\mu}{}_{\alpha}\gamma ^{\nu}{}_{\beta}\gamma ^{\gamma}{}_{\rho}\nabla
	_{\mu}(D _{\nu}v^\rho)\,.
\end{equation}
Once again for the second derivative:
\begin{equation}
	D _{\alpha}D _{\beta}v^\gamma = \gamma ^{\mu}{}_{\alpha}\gamma
	^{\nu}{}_{\beta}\gamma ^{\gamma}{}_{\rho}\nabla _{\mu}(\gamma
	^{\sigma}{}_{\nu}\gamma ^{\rho}{}_{\lambda}\nabla _{\sigma} v^\lambda)\,.
\end{equation}
We can further develop the above formula using the dependence $\nabla_\mu \gamma
^{\sigma}{}_{\nu}= \nabla _{\mu}(\delta ^{\sigma}{}_{\nu}+n^\sigma n _{\nu}) =
\nabla _{\mu}n^\sigma n _{\nu}+n^\sigma \nabla _{\mu}n_\nu$, and because $\gamma
^{\nu}{}_{\beta}n _{\nu}=0$, we get:
\begin{equation}
	\label{gauss}
	\begin{aligned}
	D_\alpha D_\beta v^\gamma =& \gamma ^{\mu}{}_{\alpha}\gamma
	^{\nu}{}_{\beta}\gamma ^{\gamma}{}_{\rho}(n^\sigma\nabla_\mu n_\nu \gamma
	^{\rho}{}_{\lambda}\nabla_\sigma v^\lambda + \gamma
	^{\sigma}{}_{\nu}\nabla_\mu n^\rho n _{\lambda}\nabla_\sigma v^\lambda +
	\gamma ^{\sigma}{}_{\nu}\gamma ^{\rho}{}_{\lambda}\nabla _{\mu}\nabla
	_{\sigma}v^\lambda) = \\
	=& \gamma ^{\mu}{}_{\alpha}\gamma ^{\nu}{}_{\beta}\gamma
	^{\gamma}{}_{\lambda}\nabla_\mu n_\nu n^\sigma \nabla_\sigma v^\lambda -
	\gamma ^{\mu}{}_{\alpha}\gamma ^{\sigma}{}_{\beta}\gamma
	^{\gamma}{}_{\rho}v^\lambda \nabla_\mu n^\rho \nabla_\sigma n_\lambda +
	\gamma ^{\mu}{}_{\alpha}\gamma ^{\sigma}{}_{\beta}\gamma
	^{\gamma}{}_{\lambda}\nabla_\mu\nabla_\sigma v^\lambda =\\
	=& -K _{\alpha \beta}\gamma ^{\gamma}{}_{\lambda}n^\sigma \nabla
	_{\sigma}v^\lambda - K ^{\gamma}{}_{\alpha}K _{\beta \lambda} v^\lambda +
	\gamma ^{\mu}{}_{\alpha}\gamma ^{\sigma}{}_{\beta}\gamma
	^{\gamma}{}_{\lambda}\nabla _{\mu}\nabla _{\sigma}v^\lambda\,,
	\end{aligned}
\end{equation}
where we used the projection operator's property $\gamma
^{\gamma}{}_{\rho}\gamma ^{\rho}{}_{\lambda}=\gamma ^{\gamma}{}_{\lambda}$ and
definition of extrinsic curvature $\gamma ^{\mu}{}_{\alpha}\gamma
^{\nu}{}_{\beta}\nabla_\mu n_\nu = - K _{\beta\alpha}$. \\
If we now exchange the order of the indices $\alpha$ and $\beta$ and subtract side by side the expressions with exchanged indices from the equation  (\ref{gauss}), we get:
\begin{equation}
	\label{gauss2}
	D_\alpha D_\beta v^\gamma - D_\beta D_\alpha v^\gamma = (K _{\alpha \mu}
	K ^{\gamma}{}_{\beta}-K _{\beta \mu}K ^{\gamma}{}_{\alpha})v^\mu + \gamma
	^{\rho}{}_{\alpha}\gamma ^{\sigma}{}_{\beta}\gamma
	^{\gamma}{}_{\lambda}(\nabla_\rho \nabla_\sigma v^\lambda - \nabla_\sigma
	\nabla_\rho v^\lambda)\,.
\end{equation}
We can use the definition of a four-dimensional Riemann tensor:
\begin{equation}
	\nabla_\rho \nabla_\sigma v^\lambda - \nabla_\sigma
	\nabla_\rho v^\lambda = \overset{4}{R}{}^{\lambda}{}_{\mu\rho\sigma}v^\mu\,,
\end{equation}
to simplify the equation (\ref{gauss2}):
\begin{equation}
	D_\alpha D_\beta v^\gamma - D_\beta D_\alpha v^\gamma = (K _{\alpha \mu}
	K ^{\gamma}{}_{\beta}-K _{\beta \mu}K ^{\gamma}{}_{\alpha})v^\mu + \gamma
	^{\rho}{}_{\alpha}\gamma ^{\sigma}{}_{\beta}\gamma
	^{\gamma}{}_{\lambda}\overset{4}{R}{}^{\lambda}{}_{\mu\rho\sigma}v^\mu\,.
\end{equation}
Let us substitute the left side of the above equation to (\ref{gauss0}):
\begin{equation}
	(K _{\alpha \mu}
	K ^{\gamma}{}_{\beta}-K _{\beta \mu}K ^{\gamma}{}_{\alpha})v^\mu + \gamma
	^{\rho}{}_{\alpha}\gamma ^{\sigma}{}_{\beta}\gamma
	^{\gamma}{}_{\lambda}\overset{4}{R}{}^{\lambda}{}_{\mu\rho\sigma}v^\mu =
	R ^{\gamma}{}_{\mu \alpha \beta}v^\mu
\end{equation}
or equivalently (because $v^\mu = \gamma ^{\mu}{}_{\sigma}v^\sigma$):
\begin{equation}
	\gamma ^{\mu}{}_{\alpha}\gamma ^{\nu}{}_{\beta}\gamma
	^{\gamma}{}_{\rho}\gamma
	^{\sigma}{}_{\lambda}\overset{4}{R}{}^{\rho}{}_{\sigma \mu \nu}v^\lambda = R
	^{\gamma}{}_{\lambda \alpha \beta}v^\lambda + (K ^{\gamma}{}_{\alpha}K
	_{\lambda \beta}-K ^{\gamma}{}_{\beta}K _{\alpha \lambda})v^\lambda\,.
\end{equation}
Both $K$ and $R$ are tangent to $\Sigma$, so the above formula is true for any vector field. We can write:
\begin{equation}
\label{gaussRelation}
	\gamma ^{\mu}{}_{\alpha}\gamma ^{\nu}{}_{\beta}\gamma
	^{\gamma}{}_{\rho}\gamma
	^{\sigma}{}_{\lambda}\overset{4}{R}{}^{\rho}{}_{\sigma \mu \nu} = R
	^{\gamma}{}_{\lambda \alpha \beta} + K ^{\gamma}{}_{\alpha}K
	_{\lambda \beta}-K ^{\gamma}{}_{\beta}K _{\alpha \lambda}\,.
\end{equation}
This result is called Gaussian formula. \\
If in the equation (\ref{gaussRelation}) we will contract the indices $\gamma$ and $\alpha$ and use the dependency $\gamma ^{\mu}{}_{\alpha} \gamma
^{\alpha}{}_{\rho}=\gamma ^{\mu}{}_{\rho} = \delta ^{\mu}{}_{\rho}+ n^\mu n
_{\rho}$, we get a relation between the Ricci tensors
$\overset{4}{R}_{\alpha\beta}$ and $R_{\alpha\beta}$:
\begin{equation}
\label{contractedGauss}
	\gamma ^{\mu}{}_{\alpha}\gamma ^{\nu}{}_{\beta}\overset{4}{R}{}_{\mu
	\nu}+\gamma _{\alpha \mu}n^\nu\gamma ^{\rho}{}_{\beta}n^\sigma
	\overset{4}{R}{} ^{\mu}{}_{\nu\rho\sigma}=R _{\alpha \beta} + K K _{\alpha
	\beta}-K _{\alpha \mu}K ^{\mu}{}_{\beta}\,,
\end{equation}
which we will call contracted Gaussian formula. \\
We can calculate the trace one more time, note that $K ^{\mu}{}_{\mu} = K
^{i}{}_{i}=K$, $K _{\mu \nu}K ^{\mu \nu}= K _{ij}K ^{ij}$ and:
\begin{equation}
	\gamma ^{\alpha \beta}\gamma _{\alpha \mu}n^\nu\gamma
	^{\rho}{}_{\beta}n^\sigma \overset{4}{R}{} ^{\mu}{}_{\nu \rho \sigma} =
	\gamma ^{\rho}{}_{\mu}n^\nu n^\sigma
	\overset{4}{R}{}^{\mu}{}_{\nu\rho\sigma}= \overset{4}{R}{}_{\mu\nu}n^\mu
	n^\nu\,.
\end{equation}
So the trace (\ref{contractedGauss}) takes the form:
\begin{equation}
	\overset{4}R+2\overset{4}{R}{} _{\mu \nu}n^\mu n^\nu = R +K^2 - K _{ij}K
	^{ij}\,.
\end{equation}
The above result is a generalization of Gauss's ``Remarkable Theorem'' (Latin: Theorema Egregium). It combines the internal curvature of $\Sigma$ (represented by Ricci
scalar $R$) with the extrinsic curvature (represented by $K^2 - K
_{ij}K ^{ij}$). The original Gaussian result concerned two-dimensional surfaces immersed in a flat Euclidean space $\mathbb{R}^3$ (for which the left hand side is zero). In addition, in the original proposition member $K^2 - K _{ij}K
^{ij}$ has the opposite sign, because the metric of ``whole'' space is Riemannian, and not (as in the case of gravity) Lorentzian.

%%%%%%%%%%%%%%%%%%%%%%%%%%%%%%%%%%%%%%%%%%%%%%%%%%%%%%%%%%%%%%%%%%%%%%%%%%%%%%%%
%                                   Codazzi                                    %
%%%%%%%%%%%%%%%%%%%%%%%%%%%%%%%%%%%%%%%%%%%%%%%%%%%%%%%%%%%%%%%%%%%%%%%%%%%%%%%%

\subsection{Codazzi's formulae}
\label{sectionCodazzi}
Let us apply the definition of Riemann tensor to the normal vector $n$ (actually to extention of $n$ to the environment of $\Sigma$):
\begin{equation}
	(\nabla_\alpha \nabla_\beta - \nabla_\beta \nabla_\alpha)n^\gamma =
	\overset{4}{R}{}^{\gamma}{}_{\mu\alpha \beta}n^\mu\,,
\end{equation}
and project the above result on $\Sigma$:
\begin{equation}
	\gamma ^{\mu}{}_{\alpha}\gamma ^{\nu}{}_{\beta}\gamma
	^{\gamma}{}_{\rho}\overset{4}{R}{} ^{\rho}{}_{\sigma \mu \nu}n^\sigma =
	\gamma ^{\mu}{}_{\alpha}\gamma ^{\nu}{}_{\beta}\gamma
	^{\gamma}{}_{\rho}(\nabla_\mu \nabla_\nu n^\rho - \nabla_\nu \nabla_\mu
	n^\rho)\,.
\end{equation}
Next, using the dependence:
\begin{equation}
	\nabla_\beta n_\alpha = -K _{\alpha \beta}-a_\alpha n_\beta\,,
\end{equation}
where $a_\alpha = n^\beta\nabla_\beta n_\alpha$, we get:
\begin{equation}
	\label{przedCodazzim}
	\begin{aligned}
	\gamma ^{\mu}{}_{\alpha}\gamma ^{\nu}{}_{\beta}\gamma
	^{\gamma}{}_{\rho}\nabla_\mu \nabla_\nu n^\rho =&\quad
	\gamma ^{\mu}{}_{\alpha}\gamma ^{\nu}{}_{\beta}\gamma
	^{\gamma}{}_{\rho}\nabla_\mu(-K ^{\rho}{}_{\nu}-a^\rho n_\nu) = \\
	=& \quad -\gamma ^{\mu}{}_{\alpha}\gamma ^{\nu}{}_{\beta}\gamma
	^{\gamma}{}_{\rho}(\nabla_\mu K ^{\rho}{}_{\nu}+\nabla_\mu a^\rho n^\nu +
	a^\rho \nabla_\mu n_\nu) = \\
	=&\quad -D_\alpha K ^{\gamma}{}_{\beta}+a^\gamma K _{\alpha \beta}\,.
	\end{aligned}
\end{equation}
We used $\gamma ^{\nu}{}_{\beta}n_\nu=0$, $\gamma
^{\gamma}{}_{\rho} a^\rho=a^\gamma$. If we exchange the order of indices
$\alpha$ and $\beta$, and the resulting expression subtract by sides from
(\ref{przedCodazzim}), then:
\begin{equation}
	\label{codazzirelation}
	\gamma ^{\gamma}{}_{\rho}n^\sigma\gamma ^{\mu}{}_{\alpha}\gamma
	^{\nu}{}_{\beta}\overset{4}{R}{}^{\rho}{}_{\sigma \mu \nu}=D_\beta K
	^{\gamma}{}_{\alpha}-D_\alpha K ^{\gamma}{}_{\beta}\,.
\end{equation}
The result (\ref{codazzirelation}) is known in the literature as Codazzi (or Codazzi--Mainardi) relation. \\
Next, we can contract the indices $\alpha$ i $\gamma$, then we get contracted Codazzi relation:
\begin{equation}
\label{contractedcodazzi}
	\gamma ^{\mu}{}_{\alpha}n^\nu\overset{4}{R}{}_{\mu\nu} = D _{\alpha}K - D
	_{\mu}K ^{\mu}{}_{\alpha}\,.
\end{equation}

\section{Killing fields in a flat space}
Below we will present a method that allows solving the equation for Killing fields and conformal Killing fields in a flat, three-dimensional Euclidean space.

%%%%%%%%%%%%%%%%%%%%%%%%%%%%%%%%%%%%%%%%%%%%%%%%%%%%%%%%%%%%%%%%%%%%%%%%%%%%%%%%
%                                ROZWIAZANIE KV                                %
%%%%%%%%%%%%%%%%%%%%%%%%%%%%%%%%%%%%%%%%%%%%%%%%%%%%%%%%%%%%%%%%%%%%%%%%%%%%%%%%

\subsection{The solution of the equation that defines the Killing vectors}
\label{KVflatspace}
We perform the calculations in Cartesian coordinates, therefore we can convert covariant derivatives into ordinary partial derivatives. Let us write the equation defining the Killing field in two ways:
\begin{equation}
	\label{kil1}
	X _{i,j} + X _{j,i} =0\quad \Rightarrow\quad (X _{i,j} + X _{j,i} =0)_{,k}
	=0\,,
\end{equation}
\begin{equation}
	\label{kil2}
	X _{i,k} + X _{k,i} =0\quad \Rightarrow\quad (X _{i,k} + X _{k,i} =0) _{,j}
	=0\,,
\end{equation}
We subtract the equation (\ref{kil2}) from the equation (\ref{kil1}) and exchange the order of partial derivatives:
\begin{equation}
	\label{kil3}
	(X _{j,k} - X _{k,j})_{,i} =0\,.
\end{equation}
Now we integrate (\ref{kil3}):
\begin{equation}
	\label{kil4}
	X _{j,k} - X _{k,j} = 2 A _{j k} \,,
\end{equation}
where $A _{j k} = - A _{k j}$ is a constant antisymmetric matrix, and add by sides the equation (\ref{kil4}) to the equation $X _{j,k} + X _{k,j}=0$,
we will get then:
\begin{equation}
	\label{kil5}
	X _{j,k} = A _{j k}\,.
\end{equation}
Integrating (\ref{kil5}), we get:
\begin{equation}
	\label{kil6}
	X _{j} = A _{jk} x^k + C _{j}  \,,
\end{equation}
where $C _{j}$ is a constant.%\\
For three-dimensional flat space, we have the freedom to choose three constants $C
_{j}$ and three non-diagonal elements of the antisymmetric matrix $A
_{ij}$, so we get six linearly independent Killing vectors.
Select the base so that three of them correspond to the three orthogonal translation generators in directions corresponding to the axes of the coordinate system, and three more correspond to the rotation generators around these axes.

%%%%%%%%%%%%%%%%%%%%%%%%%%%%%%%%%%%%%%%%%%%%%%%%%%%%%%%%%%%%%%%%%%%%%%%%%%%%%%%%
%                               ROZWIAZANIE CKV                                %
%%%%%%%%%%%%%%%%%%%%%%%%%%%%%%%%%%%%%%%%%%%%%%%%%%%%%%%%%%%%%%%%%%%%%%%%%%%%%%%%

\subsection{The solution of the equation defining conformal Killing vectors}
\label{CKVflatspace}
Consider a flat three-dimensional space, the equation for a conformal Killing field can be solved in the same way as in the subsection
\ref{KVflatspace}. Let us start with the equation:
\begin{equation}
	\label{ckil1}
	X _{i,j} + X _{j,i} = \lambda \eta _{i j} \,.
\end{equation}
Then differentiate it and take a linear combination like in (\ref{kil3}):
\begin{equation}
	\label{ckil2}
	(X _{i,j} - X _{j, i})_{,k} = \eta _{ki} \lambda _{,j} - \eta _{jk} \lambda
	_{,i} \,.
\end{equation}
Let us do the integration:
\begin{equation}
	\label{ckil3}
	X _{i , j} - X _{j,i} = \int (\lambda _{,j} dx _{i} - \lambda _{,i} dx _{j}) +
	2 A _{ij} \,,
\end{equation}
where $A _{ab}$ is a constant, antisymmetric matrix. Add (\ref{ckil1}) to
(\ref{ckil3}) and do the integration again:
\begin{equation}
	\label{ckil4}
	X _{j} = C _{j} + A _{ij}x^i + \frac{1}{2}\int \lambda dx_j + \frac{1}{2}\int
	dx_i\int(\lambda _{,i} dx _{j} - \lambda _{,j} dx _{i} )\,,
\end{equation}
where $C _{j}$ is a constant vector. Note that for $\lambda =0$ the first two
elements reproduce the solution of the Killing equation. To obtain conformal Killing fields, we still need to find the function $\lambda$, for this purpose we contract the indices $j$ and $k$ in the equation (\ref{ckil2}):
\begin{equation}
	\begin{aligned}
		&\quad\eta ^{j k} (X _{i,j k} - X _{j, i k} ) = \eta ^{j k} (\eta _{k i} \lambda
		_{,j} - \eta _{j k} \lambda _{,i}) & \Rightarrow\\
		\Rightarrow &\quad \Delta X _{i} - X ^{k} {}_{,ki} = \lambda _{,i} -3\lambda
		_{,i}  = -2 \lambda _{,i} & \Rightarrow \\
		\Rightarrow & \quad \Delta X _{i}  = -\frac{1}{2}\lambda _{,i} \,,
	\end{aligned}
\end{equation}
where in the last step we used the equation (\ref{ckillambda}) for $n=3$.
We can differentiate the above result and create a symmetric linear combination:
\begin{equation}
	\Delta (X _{i,j}  + X _{j,i}) = 2 \cdot \left( -\frac{1}{2} \right)\lambda
	_{,ij} \,.
\end{equation}
Note that on the left hand side in Laplacian there is the left hand side of the equation for conformal Killing fields:
\begin{equation}
	\label{ckil6}
	\eta _{ij}\Delta \lambda = - \lambda _{,ij}  \,.
\end{equation}
Let us make the contraction (\ref{ckil6}) in the indices $i$ and $j$, we get:
\begin{equation}
	3\Delta\lambda = -\Delta\lambda\quad \Rightarrow\quad \Delta\lambda=0\,.
\end{equation}
The above result can be inserted into the equation (\ref{ckil6}). We get that
$\lambda$ is at most linear in Cartesian coordinates $x^i$, so it can be written as:
\begin{equation}
	\lambda(x) = -2\alpha + 4\beta_i x^i \,,
\end{equation}
where ($\alpha$, $\beta_i$) are constant. Let us return to the equation
(\ref{ckil4}) and do the integration:
\begin{equation}
	X _{j} = C _{j} + A _{ij} x^i - \alpha x _{j} + \beta_i (2 x _{j} x ^{i} -
	\delta_j{}^i r^2)  \,.
\end{equation}
As in the case of Killing vectors, we have the freedom to choose three constants
$C_j$, three non-diagonal elements of the antisymmetric matrix $A _{i j}$
plus, in addition, one constant $\alpha$ and the three coordinates of the vector
$\beta_i$. Therefore, for the flat (or conformally flat) three-dimensional Euclidean space we have ten linearly independent conformal Killing fields. Select the base so that three of them correspond to the translation generators
$\mathcal{T}_{\mathbf k}$, the next three to the rotation generators $\mathcal{R}_{\mathbf k}$, the other four fields correspond to the scaling generator $\mathcal{S}$ and the three generators of proper conformal transformations
$\mathcal{K}_{\mathbf k}$. In the Cartesian coordinate system, the formulae take the following form:
\[
\mathcal{T}_k=\frac{\partial}{\partial x^k}\,,
\]
\[
\mathcal{R}_k=\varepsilon_k{}^{ij}x_i\frac{\partial}{\partial x^j}\,,
\]
\[
\mathcal{S}=x^k \frac{\partial}{\partial x^k}\,,
\]
\[
\mathcal{K}_k=x_k\mathcal{S}-\frac{1}{2}r^2\frac{\partial}{\partial x^k}\,,
\]
where $r^2=x^2+y^2+z^2$.
Different construction of CKV is presented in \cite{Czajka}.

\section{Completion of calculations for a fully charged solution}
\label{uzupelnienie}
Field components with spin-2 for a charged solution in Cartesian coordinates:
\begin{equation}
	W_{0ij0}=\frac{3m}{r^3}n_i n_j+\frac{15 \mathbf{k}}{r^4}n_i n_j -
	\frac{3}{r^4}(k_jn_i + k_i n_j)
	-\frac{\eta_{ij}}{r^3}(m+\frac{3\mathbf{k}}{r})\,,
\end{equation}
\begin{equation}
	\begin{aligned}
		W_{ijkl}=&\frac{3}{r^3}(m+\frac{3\mathbf{k}}{r})(n_i n_l \eta_{jk} - n_i n_k
		\eta_{jl} + n_j n_k \eta_{il} - n_j n_l \eta_{ik}) + \\
		+& \frac{2}{r^3}(m+\frac{3\mathbf{k}}{r})(\eta_{ik} \eta_{jl} -
		\eta_{il}\eta_{jk}) + \\
		+& \frac{3}{r^3}\varepsilon_{mkl}n^m n^p k_{,h}(\varepsilon_{pi}{}^h n_j -
		\varepsilon_{pj}{}^h n_i) + \\
		+& \frac{3}{r^3}\varepsilon_{mij}n^m n^p k_{,h}(\varepsilon_{pk}{}^h n_l -
		\varepsilon_{pl}{}^h n_k)\,,
	\end{aligned}
\end{equation}
\begin{equation}
	\begin{aligned}
		W_{0ijk}=& \frac{3}{r^3}((p_j-\mathbf{p} n_j) n_i n_k - (p_k-\mathbf{p}
		n_k)n_i n_j) + \\
		+&\frac{3}{r^4}( n_i n_k n^m \varepsilon_{mj}{}^l s_l - n_i n_j n^m
		\varepsilon_{mk}{}^l s_l) + \\
		+&\frac{3}{r^4} n^m \mathbf{s}(3\varepsilon_{mjk}n_i +\varepsilon_{mik} n_j +
		\varepsilon_{mji} n_k) + \\
		-&\frac{3}{r^4}\varepsilon_{mjk}n^m s_i + \\
		+&\frac{3}{r^3}\varepsilon_{mjk} n^m n^h \varepsilon_{hi}{}^l p_l\,.
	\end{aligned}
\end{equation}
Contractions of the electrical and magnetic parts with conformal Killing vectors:
	\begin{equation}
	\mathcal{T}_k=\frac{\partial}{\partial x^k}\,,
	\end{equation}
	\begin{equation}
	\mathcal{S}=x^k \frac{\partial}{\partial x^k}\,,
	\end{equation}
	\begin{equation}
	\mathcal{R}_k=\varepsilon_k{}^{ij}x_i\frac{\partial}{\partial x^j}\,,
	\end{equation}
	\begin{equation}
	\mathcal{K}_k=x_k\mathcal{S}-\frac{1}{2}r^2\frac{\partial}{\partial x^k}\,,
	\end{equation}
	\begin{equation}
		\begin{aligned}
			E _{ik}\mathcal{T}^k_j &= -\frac{\eta_{ij}}{2}(\frac{3\mathbf{w}}{r^2} +
		\frac{2m}{r^3} + \frac{6\mathbf{k}}{r^4}) - \frac{3}{r^2}n^k
		\mathbf{d}_{,l}(\varepsilon_{kj}{}^ln_i+\varepsilon_{ki}{}^ln_j) +
		\\&+\frac{3}{2r^2}(n_iw_j + n_j w_i) - \frac{3}{r^4}(n_ik_j+n_jk_i) +\\ &-n_in_j
		(-\frac{3\mathbf{w}}{2r^2} - \frac{3m}{r^3} - \frac{15\mathbf{k}}{r^4})\,,
		\end{aligned}
	\end{equation}
	\begin{equation}
		\begin{aligned}
		E^i{}_j\mathcal{R}_k{}^j=-\varepsilon_k{}^{li}n_l(\frac{3\mathbf{w}}{2r} +
		\frac{m}{r^2}+\frac{3\mathbf{k}}{r^3})-\frac{3}{r}n^i\mathbf{d}_{,k}-
		\varepsilon_k{}^{mj}x_m(\frac{3}{r^4}n^ik_j-\frac{3}{2r^2}n^iw_j)\,,
		\end{aligned}
	\end{equation}
	\begin{equation}
		\begin{aligned}
			E_{ij}\mathcal{K}^j_k &=\frac{m}{2r}(n_in_k+\eta_{ik})-d^l\frac{3}{2r}n^m
		(\varepsilon_{mil}n_k-\varepsilon_{mkl}n_i)+\\
		&-k^l\frac{3}{2r^2}(-n_ln_kn_i+n_k\eta_{li}-n_l\eta_{ik}-n_i\eta_{lk})+\\
		&-w^l\frac{3}{4}(-n_in_kn_l-n_k\eta_{li}+n_i\eta_{kl}-n_l\eta_{ik})\,,
		\end{aligned}
	\end{equation}
	\begin{equation}
		\begin{aligned}
		E^i{}_j S^j =
		n^i(\frac{2m}{r^2}+\frac{9\mathbf{k}}{r^3})-\frac{3}{r}n^k\mathbf{d}_{,l}
		\varepsilon_k{}^{il}+\frac{3}{2r}(n^i\mathbf{w}+w^i)-\frac{3}{r^3}k^i\,,
		\end{aligned}
	\end{equation}

	\begin{equation}
		\begin{aligned}
			B _{ik}\mathcal{T}^k_j &= -\frac{\eta_{ij}}{2}(\frac{3\mathbf{q}}{r^2} +
		\frac{2b}{r^3} + \frac{6\mathbf{s}}{r^4}) + \frac{3}{r^2}n^k
		\mathbf{p}_{,l}(\varepsilon_{kj}{}^ln_i+\varepsilon_{ki}{}^ln_j) +
		\\&+\frac{3}{2r^2}(n_iq_j + n_j q_i) - \frac{3}{r^4}(n_is_j+n_js_i) +\\ &-n_in_j
		(-\frac{3\mathbf{q}}{2r^2} - \frac{3b}{r^3} - \frac{15\mathbf{s}}{r^4})\,,
		\end{aligned}
	\end{equation}
	\begin{equation}
		\begin{aligned}
		B^i{}_j\mathcal{R}_k{}^j=\varepsilon_k{}^{li}n_l(-\frac{3\mathbf{q}}{2r} -
		\frac{b}{r^2}-\frac{3\mathbf{s}}{r^3})+\frac{3}{r}n^i\mathbf{p}_{,k}+
		\varepsilon_k{}^{mj}x_m(-\frac{3}{r^4}n^is_j+\frac{3}{2r^2}n^iq_j)\,,
		\end{aligned}
	\end{equation}
	\begin{equation}
		\begin{aligned}
			B_{ij}\mathcal{K}^j_k &=\frac{b}{2r}(n_in_k+\eta_{ik})+p^l\frac{3}{2r}n^m
		(\varepsilon_{mil}n_k-\varepsilon_{mkl}n_i)+\\
		&-s^l\frac{3}{2r^2}(-n_ln_kn_i+n_k\eta_{li}-n_l\eta_{ik}-n_i\eta_{lk})+\\
		&-q^l\frac{3}{4}(-n_in_kn_l-n_k\eta_{li}+n_i\eta_{kl}-n_l\eta_{ik})\,,
		\end{aligned}
	\end{equation}

	\begin{equation}
		\begin{aligned}
		B^i{}_j S^j =
		n^i(\frac{2b}{r^2}+\frac{9\mathbf{s}}{r^3})+\frac{3}{r}n^k\mathbf{p}_{,l}
		\varepsilon_k{}^{il}+\frac{3}{2r}(n^i\mathbf{q}+q^i)-\frac{3}{r^3}s^i\,.
		\end{aligned}
	\end{equation}
	After contraction with the normal vector:
\begin{equation}
	\label{dodatkowezN1}
	E_{ij}\mathcal{T}^j_kn^i=\frac{2m}{r^3}n_k-\frac{3}{r^3}d^ln^m\varepsilon_{mkl}
	-\frac{3}{r^4}k_k+\frac{9}{r^4}k^ln_ln_k+\frac{3}{2r^2}(w_k+w^ln_ln_k)\,,
\end{equation}
\begin{equation}
	E_{ij}\mathcal{R}^j_kn^i=-\frac{3}{r^2}(d^k-d^ln_ln^k)-\frac{3}{r^3}k^ln_p
	\varepsilon^p{}_{lk}+\frac{3}{2r}w^ln_p\varepsilon^p{}_{lk}\,,
\end{equation}
\begin{equation}
	E_{ij}\mathcal{K}^j_kn^i=\frac{m}{r}n_k+\frac{3}{2r}d^ln^m\varepsilon_{mkl}+
	\frac{3}{2r^2}k^l(n_ln_k+\eta_{kl})-\frac{3}{4}w^l(\eta_{kl}-3n_kn_l)\,,
\end{equation}
\begin{equation}
	E_{ij}\mathcal{S}^jn^i=\frac{2m}{r^2}+\frac{6}{r^3}k_ln^l+\frac{3}{r}w_ln^l\,,
\end{equation}
\begin{equation}
	B_{ij}\mathcal{T}^j_kn^i=\frac{2b}{r^3}n_k+\frac{3}{r^3}p^ln^m\varepsilon_{mkl}
	-\frac{3}{r^4}s_k+\frac{9}{r^4}s^ln_ln_k+\frac{3}{2r^2}(q_k+q^ln_ln_k)\,,
\end{equation}
\begin{equation}
	B_{ij}\mathcal{R}^j_kn^i=\frac{3}{r^2}(p^k-p^ln_ln^k)-\frac{3}{r^3}s^ln_p
	\varepsilon^p{}_{lk}+\frac{3}{2r}q^ln_p\varepsilon^p{}_{lk}\,,
\end{equation}
\begin{equation}
	B_{ij}\mathcal{K}^j_kn^i=\frac{b}{r}n_k-\frac{3}{2r}p^ln^m\varepsilon_{mkl}+
	\frac{3}{2r^2}s^l(n_ln_k+\eta_{kl})-\frac{3}{4}q^l(\eta_{kl}-3n_kn_l)\,,
\end{equation}
\begin{equation}
	\label{dodatkowezN2}
	B_{ij}\mathcal{S}^jn^i=\frac{2b}{r^2}+\frac{6}{r^3}s_ln^l+\frac{3}{r}q_ln^l\,.
\end{equation}
Integrating the expressions (\ref{dodatkowezN1})--(\ref{dodatkowezN2}) on the sphere
we get the results (\ref{firstintegral})--(\ref{lastintegral}).

\section{Dynamical equations for a fully charged solution}
Below is a direct calculation leading to dynamic equations for charges in the ``fully charged solution''.\\
To simplify the formulae, let us define four auxiliary objects:
\begin{equation}
	A_{ij}:=\frac{1}{r^3}(\eta_{ij}-3n_in_j)=-\left(\frac{1}{r}\right)_{,ij}\,,
\end{equation}
\begin{equation}
	B_{ijk}:=\left(\frac{1}{r}\right)_{,ijk}\,,
\end{equation}
\begin{equation}
	C_{ijl}:=\frac{3}{r^3}n^k(\varepsilon_{kjl}n_i + \varepsilon_{kil}n_j)\,,
\end{equation}
\begin{equation}
	D_{ijl}:=\frac{3}{2r^2}(\eta_{ij}n_l-n_i\eta_{jl}-n_j\eta_{il}-n_in_jn_l)\,.
\end{equation}
With the above definitions, the expressions for electrical and magnetic parts are as follows:
\begin{equation}
	E_{ij}=-mA_{ij}-k^lB_{ijl}-d^lC_{ijl}-w^lD_{ijl}\,,
\end{equation}
\begin{equation}
	B _{ij}=-bA _{ij}-s^lB _{ijl}+p^lC _{ijl}-q^lD _{ijl}\,.
\end{equation}
Let us calculate derivatives:
\begin{equation}
	A_{ij,l} = -\left(\frac{1}{r}\right)_{,ijl}\,,
\end{equation}
\begin{equation}
	B_{ijk,l}=\left(\frac{1}{r}\right)_{,ijkl}\,,
\end{equation}

\begin{equation}
	 C_{ijl,p}=\frac{3}{r^4}[\varepsilon_{kjl}(\eta_{ip}n^k-5n_in^kn_p)+\varepsilon_{kil}
	(\eta_{jp}n^k-5n_jn^kn_p) +\varepsilon_{pjl}n_i + \varepsilon_{pil}n_j]\,,
\end{equation}
\begin{equation}
	\begin{aligned}
		D_{ijl,p}= \frac{3}{2r^3}&[\eta_{ij}\eta_{lp}-\eta_{jl}\eta_{ip}-\eta_{il}
	\eta_{jp}-3n_p(\eta_{ij}n_l-\eta_{jl}n_i-\eta_{il}n_j) \\
	&-\eta_{ip}n_jn_l- \eta_{jp}n_in_l-\eta_{lp}n_in_j+5n_in_jn_ln_p]\,.
	\end{aligned}
\end{equation}
Next, let us calculate the contractions with the antisymmetric tensor:
\begin{equation}
	\varepsilon_m{}^{pj}\partial_pA_{ij}=0\,,
\end{equation}
\begin{equation}
	\varepsilon_m{}^{pj}\partial_pB_{ijl}=0\,,
\end{equation}
\begin{equation}
	\varepsilon_m{}^{pj}\partial_pC_{ijl}=-\left(\frac{1}{r}\right)_{,ilm}=-B_{ilm}\,,
\end{equation}
\begin{equation}
	 \varepsilon_m{}^{pj}\partial_pD_{ijl}=\frac{3}{2r^3}[2\varepsilon_{mli}+2n_jn_l\varepsilon_{mi}{}^j-4n_jn_i\varepsilon_{ml}{}^j]\,.
\end{equation}
\begin{lemat}
	The following equality is true:
	\begin{equation}
		\varepsilon_m{}^{pj}\partial_pD_{ijl}=-C_{mil}\,.
	\end{equation}
\end{lemat}
\noindent \textit{Proof}:\;
Equivalent equation:
\begin{equation}
	 \varepsilon_{mli}+n^jn_l\varepsilon_{mij}-2n^jn_i\varepsilon_{mlj}=-n^j(\varepsilon_{jil}n_m+\varepsilon_{jml}n_i)\,.
\end{equation}
Equivalently (we use $n^2=1$ and exclude one $n$ before the parenthesis):
\begin{equation}
	 n^j(n_j\varepsilon_{mli}+n_l\varepsilon_{mij}-2n_i\varepsilon_{mlj})=-n^j(\varepsilon_{jil}n_m+\varepsilon_{jml}n_i)\,.
\end{equation}
Equivalently, after moving everything to one side and pulling out $n^j$:
\begin{equation}
	 n_j\varepsilon_{mli}+n_l\varepsilon_{mij}-n_i\varepsilon_{mlj}+\varepsilon_{jil}n_m=0\,.
\end{equation}
Now notice that the left hand side is antisymmetric in all 6 possible pairs of the indices $j,i,l,m$. Because it has four indices, and the space dimension is three, the tensor on the left hand side must be zero. \qed

\noindent Let us summarize the results:
\begin{equation}
	\varepsilon_m{}^{pj}\partial_pA_{ij}=0\,,
\end{equation}
\begin{equation}
	\varepsilon_m{}^{pj}\partial_pB_{ijl}=0\,,
\end{equation}
\begin{equation}
	\varepsilon_m{}^{pj}\partial_pC_{ijl}=-B_{mil}\,,
\end{equation}
\begin{equation}
	\varepsilon_m{}^{pj}\partial_pD_{ijl}=-C_{mil}\,.
\end{equation}
We will use the equation (\ref{dynE}):
\begin{equation}
	\label{ladE}
	-\dot{m}A _{ij}-\dot{k}^lB _{ijl}-\dot{d}^lC _{ijl}-\dot{w}^lD _{ijl} =
	-p^lB _{ijl} +q^lC _{ijl}\,.
\end{equation}
Analogously for the magnetic part of the dynamic equation
(\ref{dynH}):
\begin{equation}
	\label{ladH}
	-\dot{b}A _{ij}-\dot{s}^lB _{ijl}+\dot{p}^lC _{ijl}-\dot{q}^lD _{ijl} = -d^lB
	_{ijl}-w^lC _{ijl}\,.
\end{equation}
Comparing the left hand and the right hand sides of the equations (\ref{ladE}) i (\ref{ladH}):
\begin{equation}
	\dot{m}=\dot{w}^l=\dot{b}=\dot{q}^l=0\,,
\end{equation}
\begin{equation}
	\dot{k}^l=p^l\,, \quad \dot{d}^l=-q^l\,, \quad \dot{s}^l=d^l\,, \quad
	\dot{p}^l=-w^l\,.
\end{equation}
The obtained results coincide with the results from the section \ref{chapterLinearized}.

\end{document}